\documentclass[aps,prd,twocolumn,groupedaddress,showpacs,nofootinbib]{revtex4-1}

\usepackage{amsmath}
\usepackage{amssymb}
\usepackage[cp1251]{inputenc}
\usepackage[T2A]{fontenc}
\usepackage[russian,english]{babel}
\usepackage{amsfonts}
\usepackage[breaklinks, colorlinks, citecolor=blue]{hyperref}
\usepackage{hhline}
\usepackage{array}

\usepackage{graphicx}

\usepackage{url}
\usepackage{epstopdf}
\begin{document}

%\preprint{}

\title{Constraints on dark matter annihilation by radio observations of M31}


\author{A.E.Egorov}
\email{astrotula@gmail.com}
%\thanks{}
%\affiliation{University of Southern California}

\author{E.Pierpaoli}
\email{pierpaol@usc.edu}
%\thanks{}
\altaffiliation{California Institute of Technology, MC 249-17, 1200 East California Blvd, Pasadena, CA 91125, USA}
\affiliation{Department of Physics and Astronomy, University of Southern California, 3620 McClintock Ave., SGM 408, Los Angeles, CA 90089, USA}


\date{\today}

\begin{abstract}
We used radio observations of the neighbour galaxy M31 in order to put constraints on dark matter particle mass and annihilation cross section. Dark matter annihilation in M31 halo produces highly energetic leptons, which emit synchrotron radiation on radio frequencies in the galactic magnetic field. We predicted expected radio fluxes for the two annihilation channels: $\chi\chi \rightarrow b\bar{b}$ and  $\chi\chi \rightarrow  \tau^+\tau^-$. We then compared them with available data on the central radio emission of M31 as observed by four radio surveys: VLSS (74 MHz), WENSS (325 MHz), NVSS (1400 MHz) and GB6 (4850 MHz). Assuming a standard NFW dark matter density profile and a conservative magnetic field distribution inside the Andromeda galaxy, we find that the thermal relic annihilation cross section or higher $\langle\sigma v\rangle \ge 3 \cdot 10^{-26}$ cm$^3$/s  are only allowed for WIMP masses greater  than $\approx 100$ GeV and $\approx 55$ GeV for annihilation into $b\bar{b}$ and $\tau^+\tau^-$ respectively. Taking into account potential uncertainties in the distributions of DM density and magnetic field, the mentioned WIMP limiting  masses  can be as low as 23 GeV for both channels, and as high as 280 and 130 GeV for annihilation into $b\bar{b}$ and $\tau^+\tau^-$ respectively. These mass values exceed the best up-to-day known constraints from Fermi gamma observations: 40 GeV and 19 GeV respectively [A.Geringer-Sameth and S.M.Koushiappas, Phys. Rev. Lett. 107, 241303 (2011)]. Precise measurements of the magnetic field in the relevant region and better reconstruction of the DM density profile of M31 will be able to reduce the uncertainties of our exclusion limits. 
\end{abstract}

%\keywords{DM}

\maketitle

\section{Introduction\label{Introduction}}

The physical nature of the dark matter (DM) is continuing to be not understood. The most probable candidate for the role of DM is  a Weakly Interacting Massive Particle (WIMP): a  supersymmetric partner of a Standard Model (SM) particle. There are three main approaches in attempts to detect WIMPs: accelerator searches, direct searches and indirect searches. Although great efforts have been dedicated already to all of these directions, the DM puzzle still seems to be very far from its final  and clear solution. 

This article is related to the last mentioned search strategy: indirect detection. The idea of indirect detection is based on the opportunity of pair annihilation of WIMPs causing  production of different highly energetic SM particles, which eventually  decay into stable particles like leptons, protons and others. These yields  may emit electromagnetic radiation through various mechanisms like synchrotron emission, Inverse Compton Scattering (ICS), etc. in different astrophysical objects. By detecting  such radiation we can infer WIMP properties. Since there are evidences of  DM existence  in any object from dwarf galaxies to largest galaxy clusters, any of these objects is potentially  a good target for indirect DM searches. Indirect searches of dark matter have been already extensively exploited considering all wavelength ranges from radio to gamma rays (see \cite{Bertone} for a review). 
%In general we expect two possible final results of such searches.
 Conventional $\Lambda$CDM model of cosmology suggests WIMP velocity averaged annihilation cross section $\langle\sigma v\rangle \approx 3 \cdot 10^{-26}$ cm$^3$/s, which yields the correct current DM abundance in the Universe in case  the DM is thermally produced (see e.g. \cite{Bertone}).  
Indirect searches may either lead to the discovery of DM with given particle annihilation cross section and mass \footnote{Note that annihilation cross section at the current epoch can be larger in more general models, than the standard thermal value even in the case of thermal production due to effects like Sommerfeld enhancement - see e.g. \cite{Sommerfeld}.}, or constrain the cross section to be below the level of a thermal relic over all plausible range of WIMP masses (from several GeV to several TeV). 
%which, in turn, will mean, that we do not understand DM phenomenon in cosmology well. 
The latter case may imply that in fact we do not understand the DM phenomenon in cosmology all that well, and a more sophisticated DM production mechanism is needed. 
%And all indirect detection activity is aiming to realize one of these two scenarios. 

Currently the most promising direction in indirect detection seems to be related to Fermi searches of the primary gamma emission produced directly by WIMP annihilation (see \cite{Fermi-dwarfs}). At the moment, these constraints are the strongest: they exclude WIMP masses smaller than about 40 GeV and 19 GeV for $b\bar{b}$ and $\tau^+\tau^-$ annihilation channels respectively for the standard thermal relic annihilation cross section. 
While the allowed region of WIMP parameters space is gradually shrinking, WIMP masses of hundreds and thousands GeV are still allowed. Indirect searches may allow further insights in the DM phenomenon. Wavelength bands ranging from the radio to the gamma rays and various astrophysical objects  may be exploited further to this aim.
%And specifically 
In this work we studied, in particular, radio observations of the closest big galaxy M31. Relativistic leptons produced by DM annihilation in its halo emit synchrotron radiation on radio frequencies due to Andromeda's magnetic field.
Given the size of the galaxy and its DM halo, the strength of its magnetic field and its close distance, large radio fluxes from DM annihilation may be expected from M31. Therefore M31's radio properties may be exploited in order to put strong upper limits on the annihilation cross section.
 Surprisingly, almost no attempts in this direction have been made in the past: in the literature only few articles can be found dedicated to indirect searches in M31. The most remarkable among them is \cite{gamma-M31}. These authors obtained some constraints on WIMP parameter space by Cherenkov ground based gamma observations. However, those observations did not have enough sensitivity to probe the relevant region of DM parameters, and since then no significant progress has been made on this object.

We computed the expected radio flux due to DM annihilation (section \ref{Computing the radio flux}), and then compared it with available data of radio observations of M31, which allowed us to put upper limits on the annihilation cross section (section \ref{Obtaining constraints on DM annihilation}). For comparison with real observations we chose all appropriate radio surveys, which cover a wide range of frequencies: VLSS (74 MHz), WENSS (325 MHz), NVSS (1400 MHz) and GB6 (4850 MHz).
 The limits obtained can be considered as conservative because  we did not make any specific assumptions about the radio emission other than
 the one  from DM in the center of M31, and allowed  for an unconstrained contribution from all other unknown backgrounds.

We
 %obtained our 
 calculated the constraints for two annihilation channels, specifically the ones annihilating  into $b\bar{b}$ and $\tau^+\tau^-$ pairs.
  %Of course, there are many more annihilation channels, but such choice is motivated by the following suggestions. 
 We chose these channels among all possibilities because, as explained e.g. in \cite{MW-ss}, $b\bar{b}$ and $\tau^+\tau^-$ nearly present the channels with the softest and hardest lepton yields respectively.  Any other case would therefore produce radio fluxes at intermediate levels with respect to these two. 
 In this sense, these can be considered as the two limiting cases.

\begin{figure}
  \centering
  \includegraphics[width=1\linewidth]{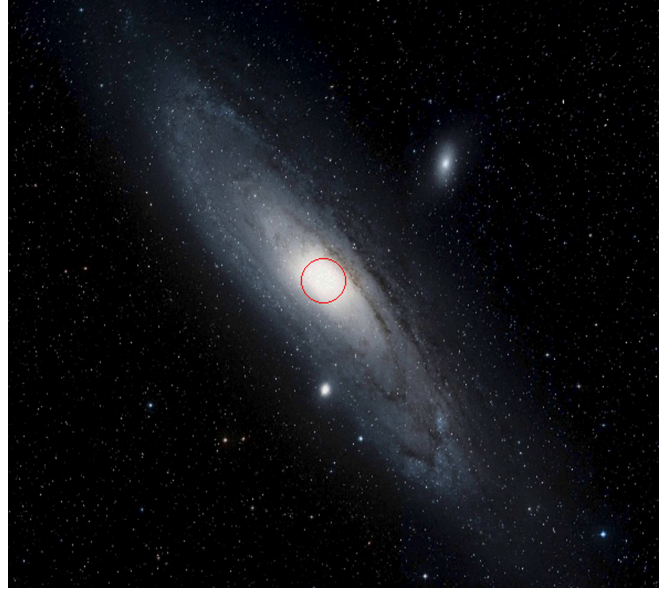}\\
  \caption{Optical image of M31 with the marked region of interest (ROI), selected for our purposes. Circle radius is 5$'$.
       }\label{ROI-optics}
 \end{figure}

In our analysis we considered only the central part of M31, and specifically the bulge area of circular shape with the angular radius $\alpha \approx 5'$ around galactic center (see fig. \ref{ROI-optics}). We chose this specific region of interest (ROI) as a target of indirect searches because of the  following considerations: {\it i)} the radio quietness of the M31 nucleus, which indicates low contamination in the radio band  by other  standard astrophysical processes; {\it ii)} the absence of any projected point source inside it; {\it iii)} sufficient halo size 
to produce a relevant signal. More details about the ROI choice will be explained below.

Through our paper we adopted the Hubble constant value $H_0 = 71$ km/(s$\cdot$Mpc), which was taken from the WMAP7 data at \url{http://lambda.gsfc.nasa.gov/}.

This paper is organized as follows: section \ref{Computing the radio flux} describes computation of the expected radio emission properties due to WIMP annihilation in M31, in section \ref{Obtaining constraints on DM annihilation} we derived the actual constraints by comparison of the predicted fluxes with observational data, and  section \ref{Summary and discussion} summarizes the results of  our work.

\section{Computing the radio flux \label{Computing the radio flux}}

\subsection{General theory \label{General theory}}
In this section we presented the procedure for the computation of the radio flux density from the center of M31.
We neglected here the potential absorption of radio emission between the source and the observer, since our estimates showed that it occurs at a negligible level (see appendix \ref{Absorption analysis}).
% The first relevant question here is potential absorption of radio emission on a way between source and observer. Galactic center presents very dense environment, that's why we needed to conduct detailed absorption analysis. Fortunately, this analysis showed negligible level of absorption. That's why further we do not include any absorption in our calculations. Details of absorption analysis are presented in Appendix 1. 
In the case of an optically transparent emitting medium, the total flux density from our ROI in M31 can be obtained just by integrating the local medium emissivity $j(\nu, \vec{r})$ over the volume of halo contained in our ROI:
\begin{equation}\label{S}
  S = \frac{\int j(\nu, \vec{r}) dV}{4 \pi d^2},
\end{equation}
where $d=785 \pm 25$ kpc is the distance between us and the Andromeda center (\cite{c=12}), $\vec{r}$ is the position vector inside the M31 halo originating in the M31 center. We disregarded here all (small) redshift effects.
%Here we are not concerned about any redshift effects due to their smallness. 
Then as a next step we needed to compute the local emission coefficient at an arbitrary position in M31 halo $j(\nu, \vec{r})$. The synchrotron emissivity of leptons produced by WIMP annihilation has the form:
\begin{widetext}
\begin{equation}\label{j}
  j(\nu,\vec{r}) = \int\limits_{m_e c^2}^{m_{\chi} c^2} \left(P_{e^+}(E_{e^+},\nu,\vec{r}) \frac{dn_{e^+}}{dE_{e^+}} dE_{e^+} + P_{e^-}(E_{e^-},\nu,\vec{r}) \frac{dn_{e^-}}{dE_{e^-}} dE_{e^-}\right) = 2\int\limits_{m_e c^2}^{m_{\chi} c^2} P_{e}(E,\nu,\vec{r}) \frac{dn_{e}}{dE} dE,
\end{equation}
\end{widetext}
where $e^+$ and $e^-$ represent positrons and electrons respectively, $P_{e}(E,\nu,\vec{r})$ is the synchrotron emission power of one lepton with energy $E$ on a frequency $\nu$ (measured in erg/(s$\cdot$Hz) in CGS), $\frac{dn_{e}}{dE}$ is the energy distribution of leptons - the number of leptons per unit volume per unit energy range. We assumed electron and positron terms in eq. (\ref{j}) to be equal to each other, which is reasonable because these both species behave similarly in all relevant aspects. According to \cite{Profumo}:
\begin{equation}\label{P}
  P_{e}(E,\nu,\vec{r}) = \int\limits_0^{\pi} d\theta' \pi \sqrt{3} \sin^2\theta' r_e m_e c \nu_0 F\left(\frac{x}{\sin \theta'}\right),
\end{equation}
where

%\begin{gather}
% r_e = \frac{e^2}{m_e c^2}, \notag \\
% x = \frac{2\nu}{3\nu_0 \gamma^2}\left(1+\left(\frac{\gamma \nu_{pl}}{\nu}\right)^2\right)^{3/2}, \notag \\
%  \nu_0(\vec{r}) = \frac{eB(\vec{r})}{2 \pi m_e c}, \label{P content} \\
%  \nu_{pl} = \sqrt{\frac{e^2 n}{\pi m_e}} - \mbox{plasma frequency of ambient plasma}, \notag \\ 
 % F(t) \approx 1.25(648)^{1/12}t^{1/3}e^{-t} - \mbox{function describing the shape of synchrotron spectrum}, \notag \\ 
 % \gamma = \frac{E}{m_e c^2} - \mbox{lepton Lorentz-factor} \notag
%\end{gather}

\begin{equation}
	\left\{
	\begin{array}{l}
	r_e = \frac{e^2}{m_e c^2},  \\ 
  x = \frac{2\nu}{3\nu_0 \gamma^2}\left(1+\left(\frac{\gamma \nu_{pl}}{\nu}\right)^2\right)^{3/2}, \\
 \nu_0(\vec{r}) = \frac{eB(\vec{r})}{2 \pi m_e c}, \label{P content} \\
  \nu_{pl} = \sqrt{\frac{e^2 n}{\pi m_e}} - \mbox{plasma frequency of ambient plasma},  \\ 
  F(t) \approx 1.25(648)^{1/12}t^{1/3}e^{-t} - \mbox{synchrotron spectrum}, \\ 
  \gamma = \frac{E}{m_e c^2} - \mbox{lepton Lorentz-factor}.
  \end{array}
  \right.
\end{equation}

\begin{figure}
  \centering
  \includegraphics[width=1\linewidth]{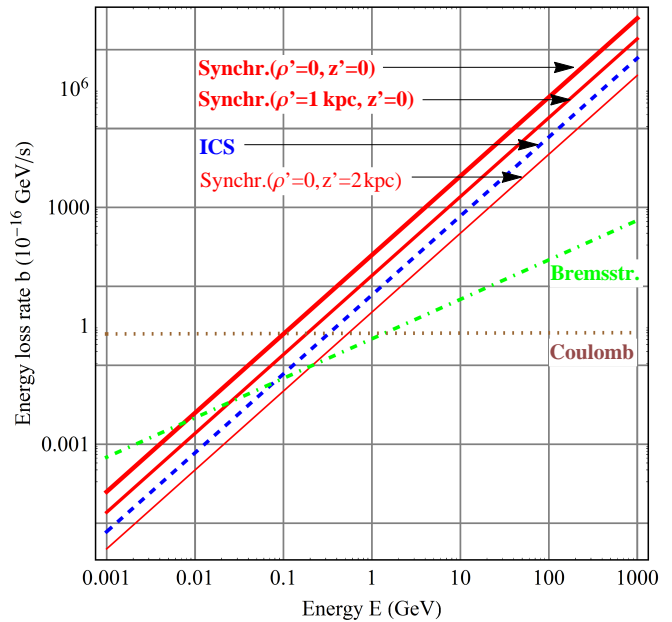}\\
  \caption{The energy dependence of  various types of lepton energy losses.  Synchrotron losses are shown for  three representative locations around the M31 center. For more details see subsections \ref{General theory} and \ref{Magnetic field}. Color version of our article is available online.}\label{Losses}
 \end{figure}

This set of formulas describes the synchrotron power of a lepton in magnetic field $B(\vec{r})$ in the presence of an ambient plasma with concentration $n$. Integration over $\theta'$ in eq. (\ref{P}) represents the averaging over all possible random angles between lepton's velocity and magnetic field. In order to proceed our computation of the radio flux expected we needed to derive the energy distribution $\frac{dn_{e}}{dE}$. For this purpose we applied the standard diffusion equation:
\begin{equation}\label{diffusion}
  \frac{\partial}{\partial t} \frac{dn_{e}}{dE} = \nabla\left(D\nabla \frac{dn_{e}}{dE}\right) + \frac{\partial}{\partial E}\left(b(E,\vec{r})\frac{dn_{e}}{dE}\right)+q_e(E,r),
\end{equation}
where $D$ is the spatial diffusion coefficient, $b(E,\vec{r})$ is the energy loss rate for a lepton through various energy dissipation mechanisms (measured in the units of energy per time), $q_e(E,r)$ is the source function - how many electrons (or positrons - but only one of these two species) are produced by WIMP annihilation per unit time per unit volume per unit energy range. We can simplify this equation assuming the stationary limit $\frac{\partial}{\partial t} = 0$,  as commonly done for diffusion of DM annihilation products in galaxies and clusters (see e.g. \cite{Borriello-MW}). Another useful simplification is the absence of the spatial diffusion of annihilation products, which will make the first term in the r.h.s. vanish. We investigated the validity of such assumption by comparing the characteristic diffusion length of newly injected leptons with the size of our ROI and concluded that it could be done without significant effect on our results (see appendix \ref{Spatial diffusion of annihilation products} for details). With these two simplifications we can easily solve the diffusion equation:
\begin{equation}\label{dn/dE}
  \frac{dn_{e}}{dE}(E,\vec{r}) = \frac{1}{b(E,\vec{r})} \int\limits_E^{m_{\chi} c^2} q_e(E',r)dE'.
\end{equation}
At this point we should specify the functions $b(E,\vec{r})$ and $q_e(E,r)$. The energy loss rate $b(E,\vec{r})$ is constituted mainly by four different cooling processes: ICS emission, synchrotron emission, bremsstrahlung and Coulomb scattering:
 \begin{equation}
 b(E,\vec{r}) = b_{ICS} + b_{sync} + b_{brem} + b_{Col}. 
 \end{equation}
 Using previous results presented in the literature (\cite{Profumo}, \cite{Bertone} and \cite{Longair}) we constructed each of these four terms respectively as follows:
\begin{equation}\label{b_ICS}
	b_{ICS} = 0.76\frac{U_{ph}}{1 \mbox{ eV$\cdot$cm}^{-3}} \left(\frac{E}{1 \mbox{ GeV}}\right)^2,
\end{equation}
here and in other loss terms numerical pre-factors like 0.76 follows from the exact analytical computation of the corresponding loss rate for a lepton. $U_{ph}$ denotes the total energy density of radiation at relevant locations. For the galactic center region it is constituted mainly by star light photons. We substituted for this quantity the fixed value $U_{ph} = 8$ eV$\cdot$cm$^{-3}$, which is quoted  in \cite{Bertone} as a characteristic value for the MW center. 
\begin{equation}\label{b_sync}
	b_{sync} = 0.025\left(\frac{B(\vec{r})}{1 \mbox{ $\mu$G}}\right)^2\left(\frac{E}{1 \mbox{ GeV}}\right)^2,
\end{equation}
\begin{equation}\label{b_brem}
	b_{brem} = 4.70\frac{n}{1 \mbox{ cm}^{-3}}\frac{E}{1 \mbox{ GeV}},
\end{equation}
where $n$ denotes the concentration of ambient plasma in cm$^{-3}$.\footnote{This loss term $b_{brem}$ was presented in another form in the work \cite{Profumo}. However, we suspect a mistake at this point in \cite{Profumo}, because our verification of this loss rate did not confirm the expression there. For this reason, we use here the expression precisely derived in \cite{Longair} (formula (6.74) there), which appears to have better justification.} For this quantity we also used the constant value $n = 0.1$ cm$^{-3}$ relying on the results of study \cite{MW-gas}, where such plasma concentration was derived as a typical average value for the Galactic center region.
\begin{equation}\label{b_Col}
	b_{Col} = 6.13\frac{n}{1 \mbox{ cm}^{-3}}\left(1+\frac{1}{75}\log\left(\frac{E}{m_e c^2 n}\right)\right).
\end{equation}
All  loss rates here are measured in the units 10$^{-16}$ GeV/s. For illustrative purposes, we showed in  fig. \ref{Losses} the energy dependences of each term in eq. (\ref{b_ICS})-(\ref{b_Col}) over the relevant range of lepton energies. We can clearly see that the synchrotron losses term depends on the location through $B(\vec{r})$. For this reason, we presented $b_{sync}$ on fig. \ref{Losses} for three relevant trial locations, which are away from the M31 center by 0, 1 and 2 kpc. More details about $B(\vec{r})$ distribution will be discussed in subsection \ref{Magnetic field}. In the most general case, the quantities $U_{ph}$ and $n$ depend on $\vec{r}$ as well, indeed. However, at the current level of accuracy of our model we made two simplifying assumptions. First, we assumed constant values for these quantities with conservative choices; and second, we adopted the values for the MW relying on high similarity between these two galaxies.    

As for the source function $q_e(E,r)$, it is computed essentially as a product of number of leptons produced by one WIMP annihilation per unit energy range $\frac{dN_e}{dE}$ and number of annihilations per unit time per unit volume: 
\begin{equation}\label{q}
  q_e(E,r) = \frac{1}{2}\left(\frac{\rho_{\scriptscriptstyle DM}(r)}{m_{\chi}}\right)^2 \langle \sigma v \rangle \frac{dN_e}{dE},
\end{equation}
where $\rho_{\scriptscriptstyle DM}(r)$ is the local DM density, $\langle \sigma v \rangle$ is thermally averaged annihilation cross section and $m_{\chi}$ is WIMP mass. Annihilation yields of leptons $\frac{dN_e}{dE}$ for two necessary channels $\chi\chi \rightarrow b\bar{b},~ \tau^+\tau^-$ were taken from the website \url{http://www.marcocirelli.net/PPPC4DMID.html}. A detailed description of this resource is presented in \cite{PPPC}. We took the pre-computed annihilation yields from this resource in their last version, which includes electroweak corrections. The  importance of these corrections for a yields computation was justified in \cite{EW}. Inclusion of electroweak corrections leads to slightly different secondary lepton yields from annihilation into primary products with different polarizations: left- or right-polarized tau-leptons, transversely- or longitudinally-polarized W-bosons, etc. But as practice showed, the difference in the final radiation flux is very minor due to this splitting. For example, the final flux for the case of M31 DM halo concentration $c_{100} = 12$, observational frequency $\nu = 74$ MHz, WIMP mass $m_{\chi} = 100$ GeV and annihilation into $\tau^+\tau^-$ differs just by $< 1\%$ for different polarizations; which is much smaller than other total flux uncertainties presented. That's why further we did not distinguish different polarizations of primary annihilation products.

For the next step we needed to specify the DM density distribution $\rho_{\scriptscriptstyle DM}(r)$, which appears in formula (\ref{q}). Next subsection is dedicated to the DM density profile derivation.

\subsection{DM density distribution \label{DM density distribution}}
As a model of DM density distribution $\rho_{\scriptscriptstyle DM}(r)$ in M31 halo we used the standard NFW profile, firstly introduced in \cite{NFW} and widely used since that due to its universality. Some other profiles potentially describe a DM distribution in galactic halos even better than NFW - this was demonstrated e.g. in \cite{Einasto} for the Einasto density profile. However, we did not use any other profiles except NFW further due to the following reasons. First, all currently available models of the DM density distribution in M31 were made in the frame of NFW profile. The second reason is that among all profiles, the NFW yields moderate emission fluxes due to DM annihilation (lower than e.g. Moore profile - see \cite{Borriello-MW}), and this makes our constraints conservative with respect to the assumptions on the adopted density profile. Also, the NFW profile  is specified by only two free parameters, while a  larger number of parameters is required by some other profiles. This choice therefore minimizes the  overall number of ``degrees of freedom'' in our model. 

Quantitatively the DM density distribution of our choice is the  following:
\begin{equation}\label{NFW}
\rho_{\scriptscriptstyle DM}(r) = \begin{cases}
\rho_{\scriptscriptstyle DM}(r=50\text{ pc}), \text{ if $0 \leqslant r < 50$ pc};\\
\frac{\rho_s}{\frac{r}{r_s}\left(1+\frac{r}{r_s}\right)^2}, \text{ if $r \geqslant 50$ pc}.
\end{cases}
\end{equation}
It presents the exact NFW profile with a minor modification, which can be called as a flat core: we do not extrapolate the density below the radial distance of $r_t = 50$ pc, inside this central region we leave DM density at the constant level of $\rho_{\scriptscriptstyle DM}(r=50\text{ pc})$. This choice is motivated by the divergence of the NFW distribution towards the  central  point of the halo, which prevents  to estimate the DM density reliably in the very central region. The NFW profile provides reliable density estimate down to  the radius $\sim$ 50 pc for a  M31--size halo (see e.g. fig. 4 in \cite{Frenk}). That's why in order to be conservative in our computation of the medium emissivity, which depends quadratically on the local DM density, we truncated NFW profile at $r = r_t = 50$ pc and left the density at smaller radii on the level of $\rho_{\scriptscriptstyle DM}(r=50\text{ pc})$. 
%This eliminates potential final flux overestimations due to overestimation of local DM density. 
We tested the dependence of our results on the specific arbitrary choice of the truncation radius  $r_t$.
This study showed that a decrease (increase) of $r_t$ by a factor two (a factor four) produce an increase (decrease) of $\approx 3\%$
($\sim 10\%$) in final fluxes, which is far below the overall level of accuracy of our model.% {\ep so which regain is dominating? can you give a qualitative  interpretation of this?  -  I didn't really get what you meant here. As for a qualitative interpretation - I'm not sure, just NFW formula works so...}.

The DM density profile eq. (\ref{NFW}) is completely defined for a specific halo by two parameters - scaling radius $r_s$ and scaling density $\rho_{s}$. Thus, we needed to estimate them for M31. These two parameters are unambiguously linked to another two halo parameters, which are more meaningful - the halo mass $M_{\Delta}$ and its concentration $c_{\Delta}$. $\Delta$ here denotes an overdensity: halo mass $M_{\Delta}$ by definition means the DM halo mass enclosed in the sphere, average density inside which is equal to $\Delta \rho_{crit}$ with $\rho_{crit} = \frac{3H_0^2}{8 \pi G} = 9.5 \cdot 10^{-30}$ g$\cdot$cm$^{-3}$ being the current critical density of the Universe. In our article we used $\Delta = 100$ due to some practical circumstances. However, the parameters of the actual DM density distribution (\ref{NFW}) do not depend indeed on a $\Delta$ choice. Halo mass and concentration are connected with the NFW profile parameters through the following relations:
\begin{equation}\label{rvrs}
  r_{\Delta} = \left(\frac{3 M_{\Delta}}{4 \pi \Delta \rho_{crit}}\right)^{1/3}, ~c_{\Delta} \equiv \frac{r_{\Delta}}{r_s}, ~M_{\Delta} = \int\limits_0^{r_{\Delta}} \rho_{\scriptscriptstyle DM}(r) 4 \pi r^2 dr,
\end{equation}
%where the first relation presents essentially the definitions of a halo mass and radius, the second - definition of a concentration, and the third one is quite obvious. 
%{\ep there is no explanation why $c_{100}$. everything above is for a Delta, not 100}.
As for the determination of the relevant parameters for M31, 
%Hence as a next step we needed to find in the literature the values of $M_{100}$ and $c_{100}$. 
we used the results of $M_{100}$ and $c_{100}$ from \cite{c=12} and \cite{c=28}. While these are in a good agreement with each other in determination of $M_{100}$ ($(1.2 \pm 0.3) \cdot 10^{12} M_{\odot}$ vs. $(0.91 \pm 0.16) \cdot 10^{12} M_{\odot}$ respectively),  they show a rather big discrepancy in the estimation of $c_{100}$ (the most probable values cited being $\approx 12$ and $\approx 28$ respectively). 
The expected radiation fluxes from these two  concentration values differ by about one order of magnitude.
%Physically concentration means a level of crowding of DM toward the center - larger concentration implies larger central density of DM and, therefore, larger expected radiation flux from DM annihilation. As practice showed, two different values of $c_{100}$ mentioned above generate significantly different expected fluxes - with difference of $\sim 10$ times. 
This is why we decided to treat these two cases separately. They can be considered as two limiting cases yielding the most conservative constraints for $c_{100} = 12$ and the most optimistic constraints for $c_{100} = 28$. 
%And the real DM density distribution lies somewhere between these two extremes.

At this point we were ready to specify the NFW profile by estimation of its two parameters $\rho_s, r_s$. For the first case $c_{100} = 12$ we took $M_{100} = 1.2 \cdot 10^{12} M_{\odot}$ from \cite{c=12} and obtained $\rho_s, r_s$ by formulas (\ref{rvrs}). For the second case $c_{100} = 28$ we did not need to calculate $\rho_s, r_s$ - they were taken
% in prepared form 
directly from the table 2 in \cite{c=28} as the best-fitting values.

Another potentially relevant question in this subsection is substructures contribution in our flux. It is well known that any DM halo contains a lot of small subhalos inside it (see e.g. \cite{Bertone}). These subhalos are very dense and numerous, that's why they are able to substantially  increase the total flux due to DM annihilation from a whole halo  - by 10-100 times (see e.g. \cite{subs}). But in our case we can neglect substructure contribution, because the main part of our expected flux comes from very central region with a size about 1 kpc. As can be seen e.g. in \cite{subs}, substructures do not survive so close to the center due to tidal disruption, and their contribution to the total expected flux would be negligible at so small radii. Thus we did not need to include subhalos in our analysis. Moreover, subhalos' presence would only increase the  overall flux, so that the constraints obtained here are to be considered conservative.
% with respect to this point as well.

Thus, at this point we have completely specified the DM density distribution in M31 halo. Now we can move to the next step - specifying the magnetic field distribution.

\subsection{Magnetic field \label{Magnetic field}}
The emission due to DM annihilation, which we hope to detect, is generated through the  synchrotron mechanism. For synchrotron emission modeling it's crucially important to know the magnetic field strength distribution in our emitting volume. A global axis symmetry of the large spiral galaxy M31 naturally suggests assumption of two-parametric magnetic field distribution $B(\rho', z')$, where $B$ depends on the radial distance from the center $\rho'$ in the galactic plane and the vertical height $z'$ above the galactic plane.

As for $B$ dependence on the vertical coordinate $z'$, it's plausible to assume exponential dependence $B \sim \exp(-|z'|/z_0')$ with some scale height $z_0'$. Such dependence is commonly used for large spiral galaxies - see e.g. \cite{M33}. In order to 
%propose 
consider a reasonable radial dependence of $B$ in the galactic plane we used the findings of 
% conclusions of the papers
 \cite{M31-MF},\cite{MW-MF}. In \cite{M31-MF} large-scale magnetic field in the disk plane was measured between radial distances of 6 and 14 kpc. These measurements yielded almost constant magnetic field over the whole mentioned annular region of $\approx 7~\mu$G, which is the expected value for galactic discs (see e.g. \cite{Beck}). As for the magnetic field strength in the inner region $\rho' < 6$ kpc, we were unable to find reliable data. Some measurements for this region were reported only in \cite{M31-center} and only for two trial locations with poor justification. Besides this information on the central field properties we can reliably assume probably only one thing - that the field grows towards the center. In such situation we made a decision to introduce the following field distribution (including the vertical dependence part):
\begin{equation}\label{B}
  B(\rho', z') = \left(B_{10}+B_s\exp\left(-\frac{\rho'}{\rho_0'}\right)\right)\exp\left(-\frac{|z'|}{z_0'}\right).
\end{equation}
Here $B_{10} = 7~\mu$G - the plateau value observed on $\rho' \approx 6 - 14$ kpc. The term $B_s\exp\left(-\frac{\rho'}{\rho_0'}\right)$ presents an exponentially decreasing central spike with the characteristic radial extent $\rho_0'$. Of course, such distribution has a non--physical plateau extending to infinity $B(\rho'=\infty, z'=0) = 7~\mu$G. But this plateau does not matter for our calculations: at  distances of tens of  kpc DM density becomes very small, and a contribution of these regions to the overall flux is tiny. The distribution in eq. (\ref{B}) can be completely specified by two free parameters: the vertical scale height $z_0'$ and the central field value $B(0,0) = B_{10}+B_s$. After specifying these two parameters, the radial scale length $\rho_0'$ becomes  automatically set by smooth connection of the central exponential spike $B_s\exp\left(-\frac{\rho'}{\rho_0'}\right)$ to the plateau $B(\rho'>6 \mbox{ kpc}, z'=0) \approx 7~\mu$G.

Now let's discuss the specification of the values of $z_0'$ and $B(0,0)$. We estimated the scale height $z_0'$ using \cite{z0} and \cite{M31-MF}. The first mentioned article outlines such a general property of galactic magnetic fields: the scale height $z_0'$ is typically about 4 times greater than the scale height of a synchrotron emission, which arises from cosmic rays in a whole galaxy volume. The last mentioned synchrotron scale height was reported in \cite{M31-MF} for M31 with a value $\approx 0.3$ kpc. This would yield $z_0' \approx 0.3\mbox{ kpc} \cdot 4 = 1.2$ kpc. Moreover, such a value for $z_0'$ is confirmed by the authors of \cite{M31-MF} on the basis of  other independent considerations: they conclude that the minimal expected $z_0'$ value is not less than $\approx 1$ kpc. 
%Based on these considerations, 
We therefore  
%accepted
chose  $z_0' = 1.2$ kpc for all our calculations. This estimate reflects a conservative choice, since lower values seem to be unrealistic and possible greater values can only increase the final expected flux and, hence, strengthen the final constraints.

As for the second necessary parameter - $B(0,0)$, the situation is much less certain. As we already mentioned, we found in the literature only one attempt to measure the central field values in M31 - \cite{M31-center}. This work reports the total field strengths for the two locations: $B(\rho' \approx 0.3\mbox{ kpc}, z' \approx 0) = 15 \pm 3~\mu$G and $B(\rho' \approx 0.9\mbox{ kpc}, z' \approx 0) = 19 \pm 3~\mu$G. These values were obtained assuming energy equipartition between cosmic ray particles and magnetic fields. 
Such assumption, however, has been shown to lead to an underestimation of the magnetic field in similar setting.  For example, 
%However, the article \cite{MW-MF} shows in detail that this simple assumption can fail to yield correct field values. 
the authors of \cite{MW-MF} explored the central field properties of the MW and derived the stringent lower limit of 50$~\mu$G on the field strength based on multi-wavelength studies of a cosmic ray emission. At the same time, they mentioned that a simple equipartition assumption implies a central field strength of only $\sim 10~\mu$G. As for the upper bound on a central field strength, \cite{MW-MF} demonstrates that there is no certainly known value for it. Values of few hundreds $\mu$G are mentioned to be absolutely possible.

Considering the possible underestimation of M31 magnetic field due to the equipartition assumption (\cite{M31-center}), and the  similarity between MW and M31,  we chose $B(0,0) = 50~\mu$G as the most probable value. However, we performed all our calculations for alternative field values as well (which are considered as less probable) in order to study the dependence of our final results on this model parameter. These alternative less realistic values were chosen to be $B(0,0) = 15~\mu$G and $B(0,0) = 300~\mu$G. They can be considered as approximate boundaries of the interval where the actual field value lies. The detailed discussion of the final results variation due to uncertainties in $B(0,0)$ will be done in the subsection \ref{Joint analysis}. 

\begin{figure}
  \centering
  \includegraphics[width=1\linewidth]{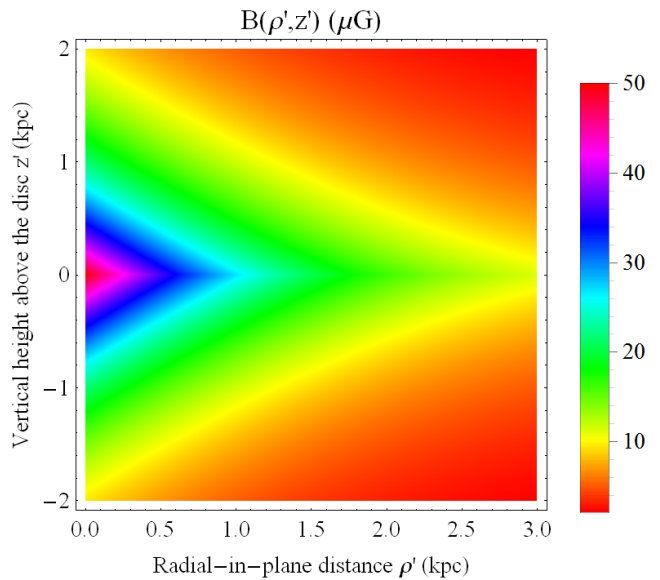}\\
  \caption{Density plot of the magnetic field distribution (\ref{B}), which was chosen for our computations. Parameter values here correspond to the most probable case with the central field strength $B(0,0)=50~\mu$G. For more details see subsection \ref{Magnetic field}. Color version of our article is available online.}\label{MF}
\end{figure}

At this point we have completely  specified magnetic field distribution. 
For illustration purposes we showed the density plot of our distribution in eq. (\ref{B}) on fig. \ref{MF} for the case of $B(0,0)=50~\mu$G.

\begin{figure*}
  \centering
  \includegraphics[width=1\linewidth]{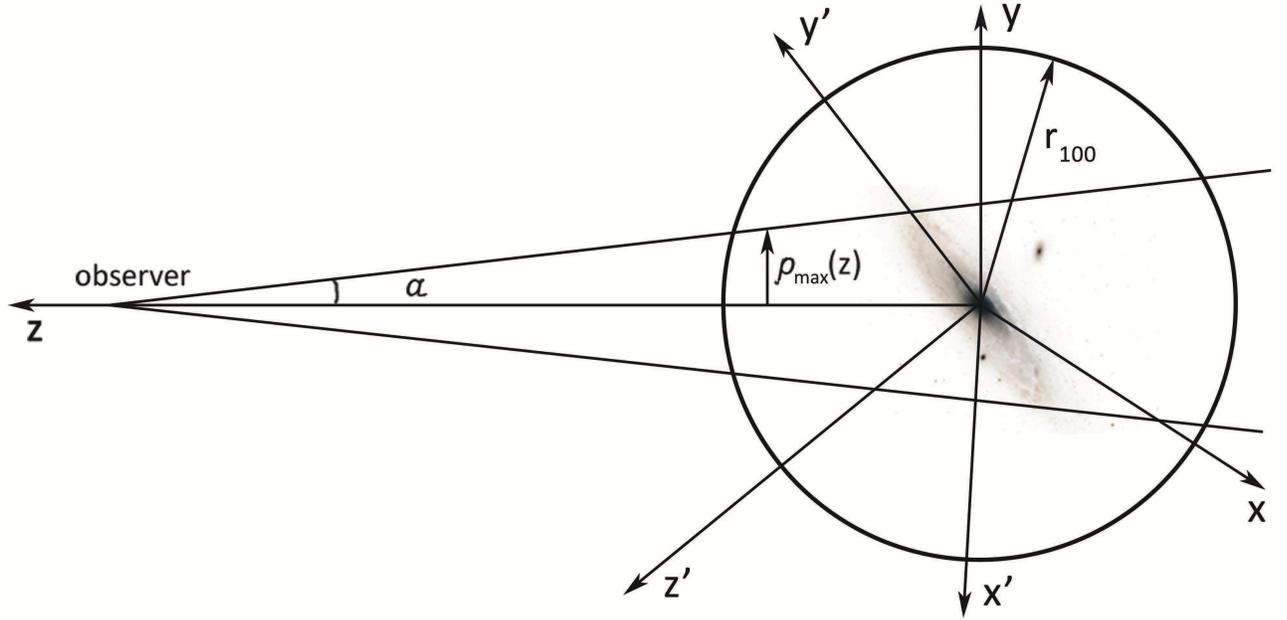}\\
  \caption{Two coordinate systems $(x,y,z)$ and $(x',y',z')$. The plane $x'y'$ is coincident with the M31 disc plane. Integration over the emitting volume is performed in $(x,y,z)$. More details are in subsection \ref{Magnetic field}.}\label{plot coordinates}
\end{figure*}

The magnetic field distribution in eq. (\ref{B}) is written in  a cylindrical coordinate system attached to M31. Axes $x'y'$ lie in the M31 disk plane. Another non trivial step is to make a transformation between coordinate system attached to M31 and "laboratory" system with $Oz$ axis pointing from the M31 center to the observer. This is necessary for the flux computation, because we are going to perform a numerical integration over the emitting volume along our line of sight, which is coincident with $Oz$ axis and not coincident with $Oz'$. 
%Essentially it's too difficult to describe emitting volume in the primed system. 
%This situation
The two coordinate systems are shown on fig. \ref{plot coordinates}. 
%That's why we made 
We performed the coordinate transformation $(\rho,\varphi,z)\longleftrightarrow(\rho',\varphi',z')$
%. For this purpose we used the following set of transformation 
using the formulas below, which were derived in \cite{coordinates} (formula (6))  for cartesian coordinates $(x,y,z)\longleftrightarrow(x',y',z')$:
\begin{equation}\label{xyz}
\begin{cases}
  x' = -x\sin{P}+y\cos{P}, \\
  y' = -x\cos{P}\cos{i}-y\sin{P}\cos{i}-z\sin{i}, \\
  z' = -x\cos{P}\sin{i}-y\sin{P}\sin{i}+z\cos{i},
\end{cases}
\end{equation}
where $P$ and $i$ represent the  position angle and axis inclination respectively, and  define the orientation of the M31 disc plane with respect to the sky plane. 
%We took values for them from
Following  \cite{c=12}, we assumed: $P=38^{\circ}, i=78^{\circ}$. From  eq. (\ref{xyz}),  using relations between cylindrical and cartesian coordinates it's easy to obtain the one-to-one correspondence $(\rho,\varphi,z)\longleftrightarrow(\rho',\varphi',z')$:
%which was actually needed:
\begin{widetext}
\begin{equation}\label{cyl}
\begin{cases}
\rho' = \left((-\rho \cos{\varphi}\sin{P} + \rho \sin{\varphi}\cos{P})^2 + (\rho \cos{\varphi}\cos{P}\cos{i} + \rho\sin{\varphi}\sin{P}\cos{i} + z\sin{i})^2\right)^{\frac{1}{2}}, \\
z' = -\rho \cos{\varphi}\cos{P}\sin{i} - \rho \sin{\varphi}\sin{P}\sin{i} + z\cos{i}.
\end{cases}
\end{equation}
\end{widetext}
%Thus we have obtained necessary transformations. Then
Finally,  we substituted eq. (\ref{cyl}) into eq. (\ref{B}) obtaining therefore  $B$ as a function of $(\rho,\varphi,z)$, which allowed to compute the final flux.
%. At this point magnetic field distribution had been completely prepared for computation of the final flux.

\subsection{Estimated fluxes \label{Estimated fluxes}}
\begin{figure*}
\begin{minipage}[h]{0.49\linewidth}
\center{\includegraphics[width=1\linewidth]{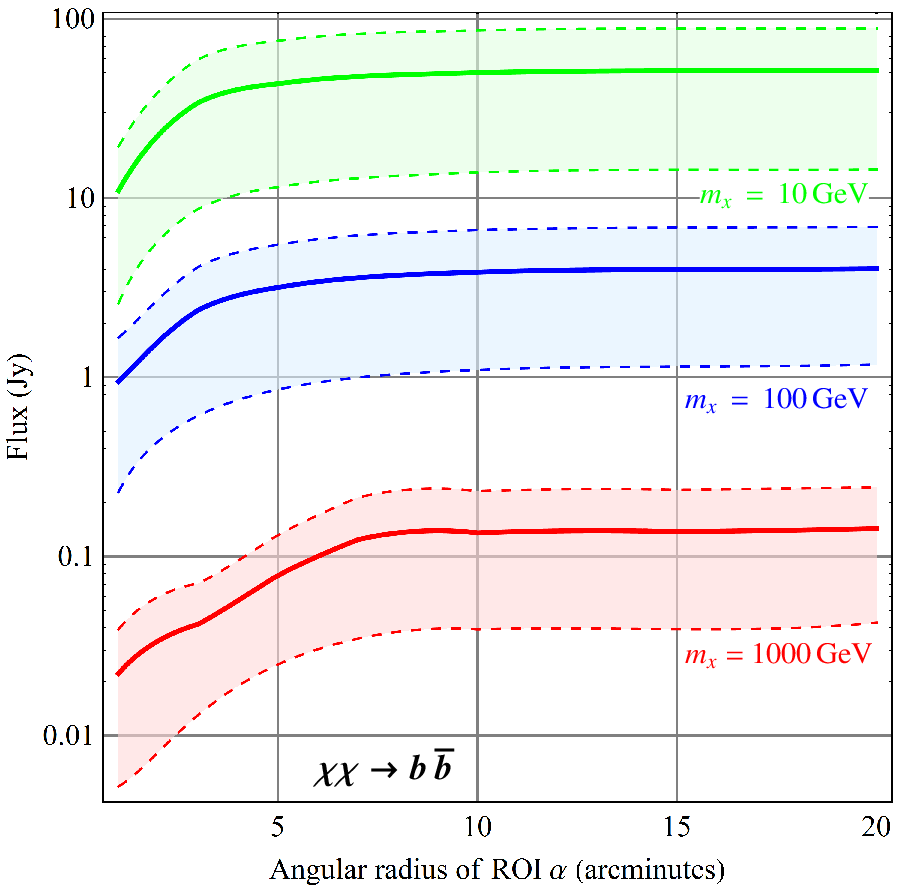} \\ }
\end{minipage}
\hfill
\begin{minipage}[h]{0.49\linewidth}
\center{\includegraphics[width=1\linewidth]{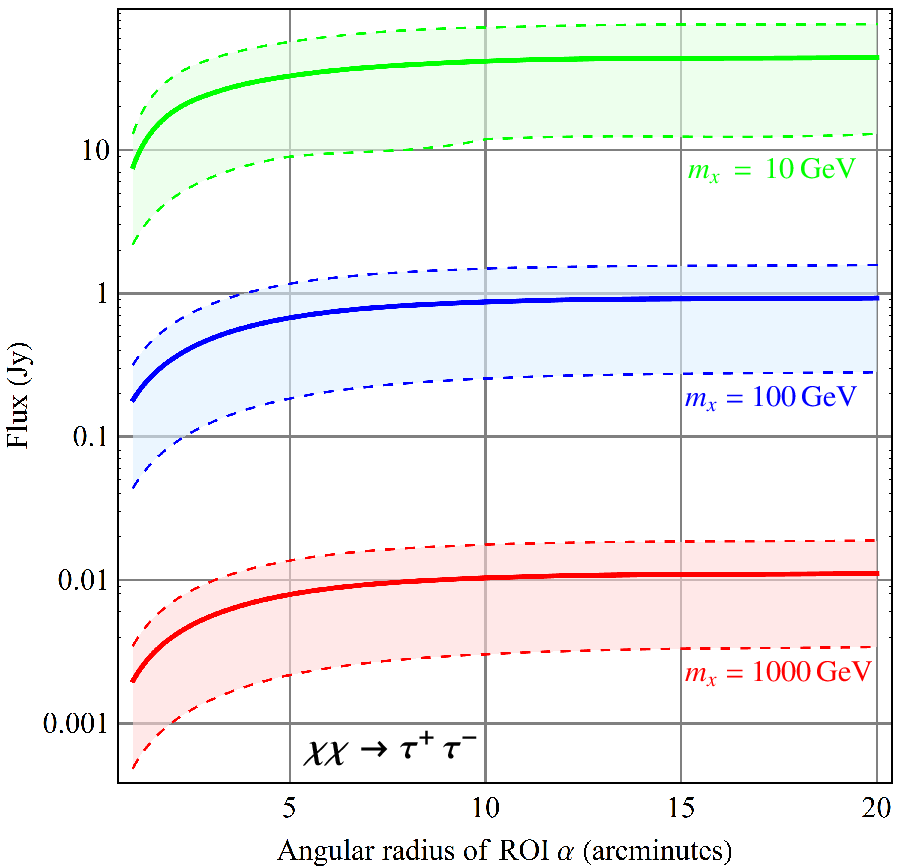} \\ }
\end{minipage}
\caption{The expected radio flux dependence on the angular radius of the ROI. The expected flux lies inside shaded regions. Upper and lower boundaries of these regions correspond to the different halo concentration values $c_{100}=28$ and $c_{100}=12$ respectively. Thick central lines present algebraic averages between corresponding limiting cases. Curves are shown for the two annihilation channels $\chi\chi \rightarrow b\bar{b},~ \tau^+\tau^-$, three WIMP masses $m_{\chi} = 10, 100$, 1000 GeV and the most probable magnetic field distribution with the central value of 50 $\mu$G. The annihilation cross section is $\left\langle \sigma v\right\rangle = 3 \cdot 10^{-26}$ cm$^3$/s, the observational frequency is $\nu$ = 74 MHz. For more details see subsection \ref{Estimated fluxes}.}
\label{S(a)}
\end{figure*}

\begin{figure*}
\begin{minipage}[h]{0.49\linewidth}
\center{\includegraphics[width=1\linewidth]{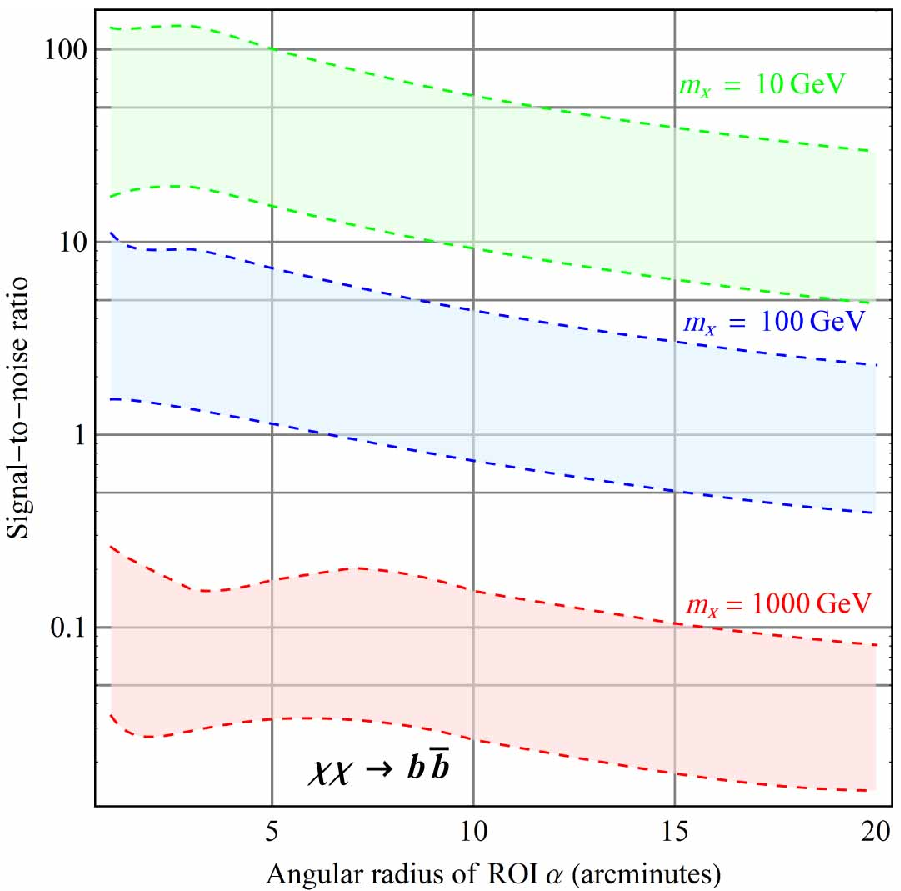} \\ }
\end{minipage}
\hfill
\begin{minipage}[h]{0.49\linewidth}
\center{\includegraphics[width=1\linewidth]{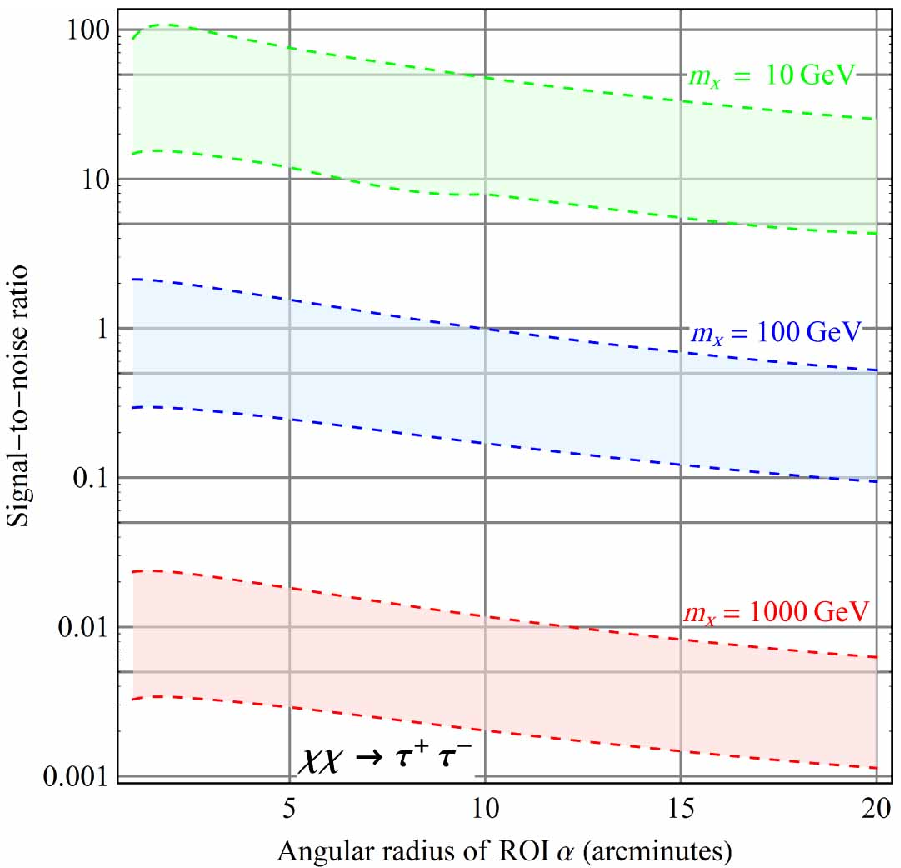} \\ }
\end{minipage}
\caption{The expected SNR dependence on the angular radius of ROI. The expected SNR lies inside shaded regions. Upper and lower boundaries of these regions correspond to different halo concentration values $c_{100}=28$ and $c_{100}=12$ respectively. Curves are shown for the two annihilation channels $\chi\chi \rightarrow b\bar{b},~ \tau^+\tau^-$, three WIMP masses $m_{\chi} = 10, 100$, 1000 GeV and the most probable magnetic field distribution with the central value of 50 $\mu$G. The annihilation cross section is $\left\langle \sigma v\right\rangle = 3 \cdot 10^{-26}$ cm$^3$/s, the observational frequency is $\nu$ = 74 MHz. For more details see subsection \ref{Estimated fluxes}.}
\label{SNR}
\end{figure*}

A final expected radio flux due to DM annihilation in M31 halo is defined by eq. (\ref{S}). In order to compute it we combined all relevant relations (\ref{j})-(\ref{P content}), (\ref{dn/dE})-(\ref{B}), (\ref{cyl}) and substituted them in eq. (\ref{S}). The integration limits for spatial coordinates can be easily figured out from the fig. \ref{plot coordinates}:
\begin{equation}\label{limits}
\begin{cases}
   0 \leqslant \rho \leqslant (d-z)\tan{\alpha}, \\
   0 \leqslant \varphi \leqslant 2\pi, \\
   -r_{100} \leqslant z \leqslant r_{100},
\end{cases}
\end{equation}
where $\alpha$ is the angular radius of our ROI. Fig. \ref{S(a)} presents the dependence of the total expected radio flux  at $\nu = 74$ MHz on the angular radius of ROI for both selected annihilation channels and three trial WIMP masses: $m_{\chi}=10, 100$ and 1000 GeV. Boundaries of every shaded region correspond to the two limiting cases of DM density distribution in the halo, which were discussed in subsection \ref{DM density distribution}: $c_{100} = 12$ and $c_{100} = 28$. Thus the actual flux is expected to lie somewhere inside shaded regions. The thick central lines present the algebraic averages between the corresponding boundary curves (fluxes). 
%These trial curves are generated for the frequency of observations $\nu = 74$ MHz. 
%We can see, that 
The flux saturates around $\alpha \approx 5'$ and does not grow significantly at larger radii.
We also computed the expected signal-to-noise ratio (SNR) for the same set of parameters and presented it on fig. \ref{SNR}. Details about noise level computations can be found in subsection \ref{Comparison with radio surveys}. These plots justify our choice of $\alpha$. Basically, in this choice we are governed by two counteracting effects. On one hand, fig. \ref{SNR} clearly suggests to use the smallest $\alpha$ possible in order to achieve the best SNR. On the other hand, as  discussed in subsections 2.2-2.3,   the accuracy of DM density and magnetic field strength estimates decreases when approaching  the central point of the halo. As a matter of fact, we know almost nothing about real distributions of these quantities inside the region of size $\alpha \approx 0.25'$. 
%That's why we can not obtain precise constraints relying on a small ROI.
 Hence, our ROI should be much larger than $\approx 0.25'$ in radius in order to encompass a  volume where our overall knowledge of parameters distribution is precise enough. Based on these considerations, we chose $\alpha = 5'$ as the optimal ROI radius. With such choice we lose little in SNR, however, we gain in overall reliability of calculations. Also, we can not increase $\alpha$ any higher than $\sim 5'$, because it will lead to non-desirable capturing of projected point sources.

\begin{figure*}
\begin{minipage}[h]{0.49\linewidth}
\center{\includegraphics[width=1\linewidth]{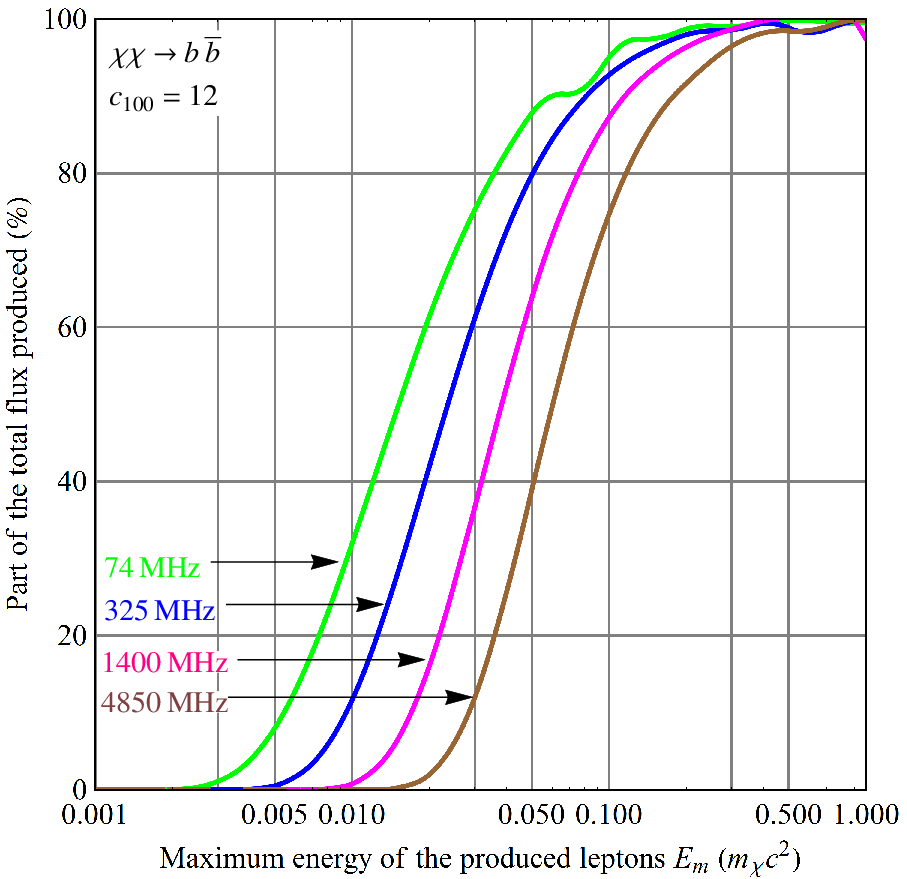} \\ }
\end{minipage}
\hfill
\begin{minipage}[h]{0.49\linewidth}
\center{\includegraphics[width=1\linewidth]{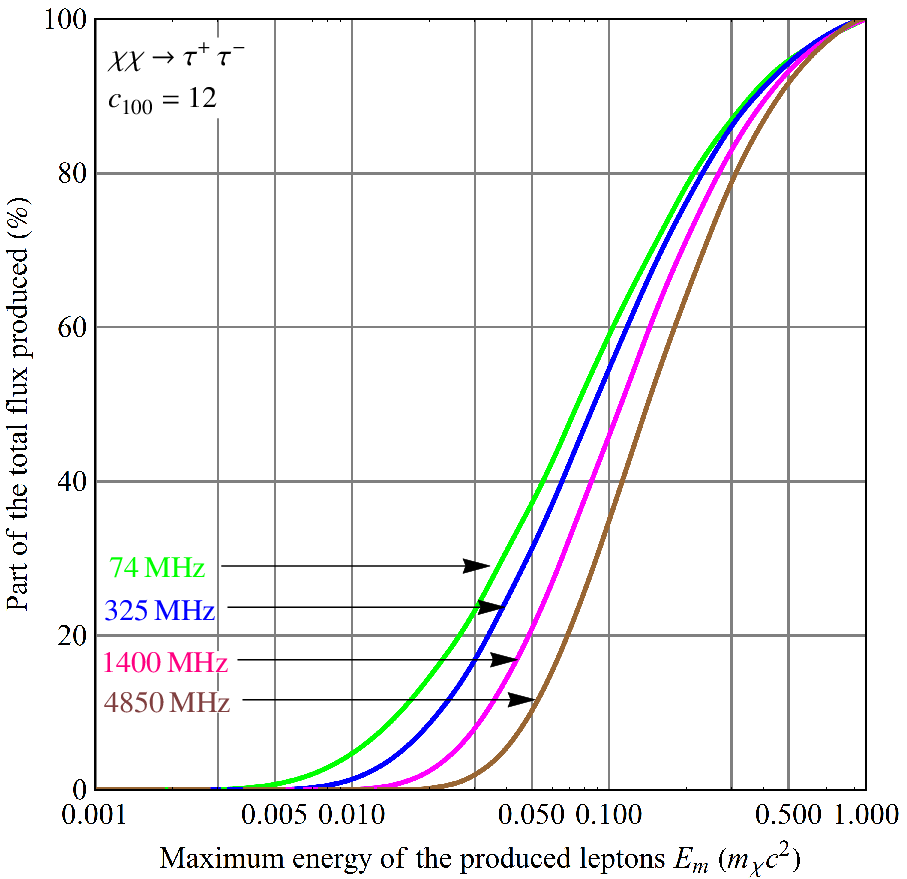} \\ }
\end{minipage}
\caption{The part of total radio flux produced by leptons with energies at production below different thresholds. These illustrations are generated for the following set of parameters: $c_{100}=12$, $\left\langle \sigma v\right\rangle = 3 \cdot 10^{-26}$ cm$^3$/s, $B(0,0)=50~\mu$G. WIMP mass is chosen to be $m_{\chi}=100$ GeV, which reflects the most relevant mass scale with respect to final results, as will be seen below. We can note that the majority of the total flux are produced by leptons with initial energies above $\sim 0.01 m_{\chi} c^2$. More details are in subsection \ref{Estimated fluxes}.}
\label{dF}
\end{figure*}

As an intermediate step of calculations, we decided to study also the contribution of leptons with different energies into the total flux. For this purpose we computed the part of the total flux produced by the leptons with initial energies at production below different thresholds $E_m$. We spanned the whole possible range of $E_m$ values from the lepton's rest energy to the WIMP's rest energy. The results are presented on fig. \ref{dF} for all four frequencies used. These plots are generated for the most relevant magnetic field model, WIMP mass $m_{\chi}=100$ GeV (relevance of this mass scale will be seen in the results section) and the $c_{100}=12$ halo model. Essentially, we can see on fig. \ref{dF} that the main contribution to the total flux ($\gtrsim 50\%$) for the all frequencies and annihilation channels comes from a relatively narrow window of initial lepton energies between $\sim 0.01 m_{\chi} c^2$ and $0.1 m_{\chi} c^2$, or 1 GeV and 10 GeV. This energy window of one order of magnitude width is considered to be narrow with respect to the whole energy range of produced leptons between $\sim 10^{-5} m_{\chi} c^2$ and $1 m_{\chi} c^2$, which spans more than 5 orders of magnitude. This implies that the most important fraction of the  lepton population for our considerations  has an energy above $\sim 1$ GeV. This fact, in turn, means that according to fig. \ref{Losses} the main energy loss mechanisms for such population would be synchrotron and ICS emission. Thus, these kinds of losses dominate in comparison with the other two - bremsstrahlung and Coloumb losses, and play a main role for final results. Meanwhile, it's also useful to note that in the case of $\tau^+\tau^-$ channel the total flux is constituted by more energetic leptons than the flux for $b\bar{b}$ channel, which is in complete agreement with the difference between these two channels mentioned in the Introduction - $\tau^+\tau^-$ channel produces harder lepton energy spectrum.  

\begin{figure*}
 \begin{minipage}[h]{0.49\linewidth}
\center{\includegraphics[width=1\linewidth]{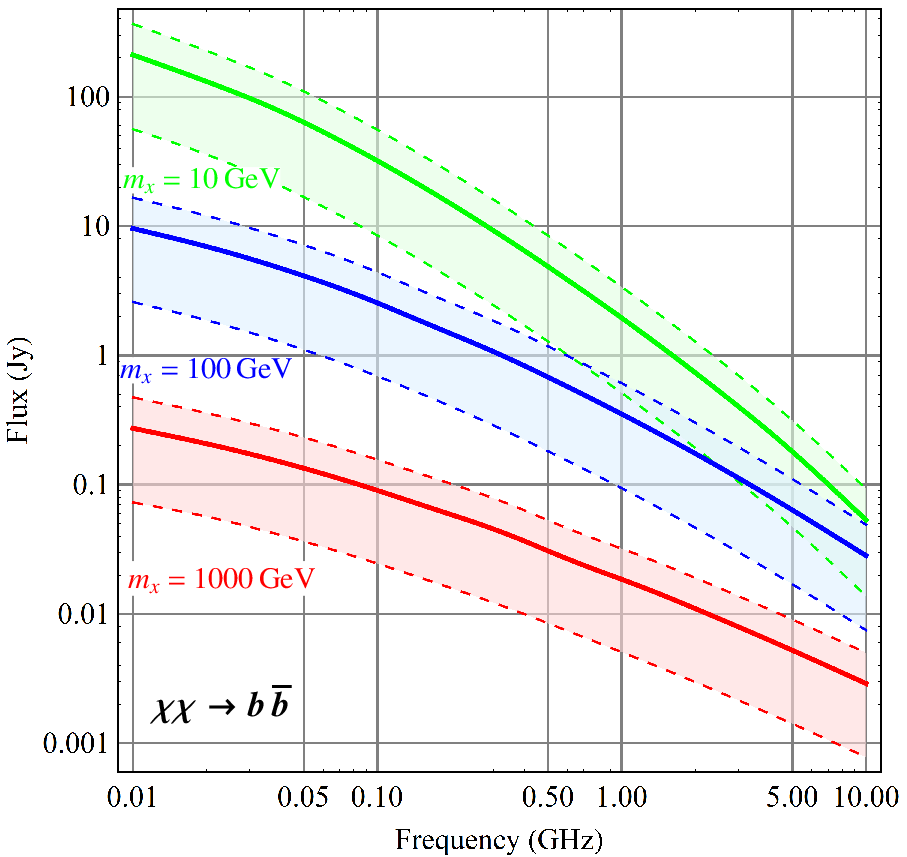} \\ }
\end{minipage}
\hfill
\begin{minipage}[h]{0.49\linewidth}
\center{\includegraphics[width=1\linewidth]{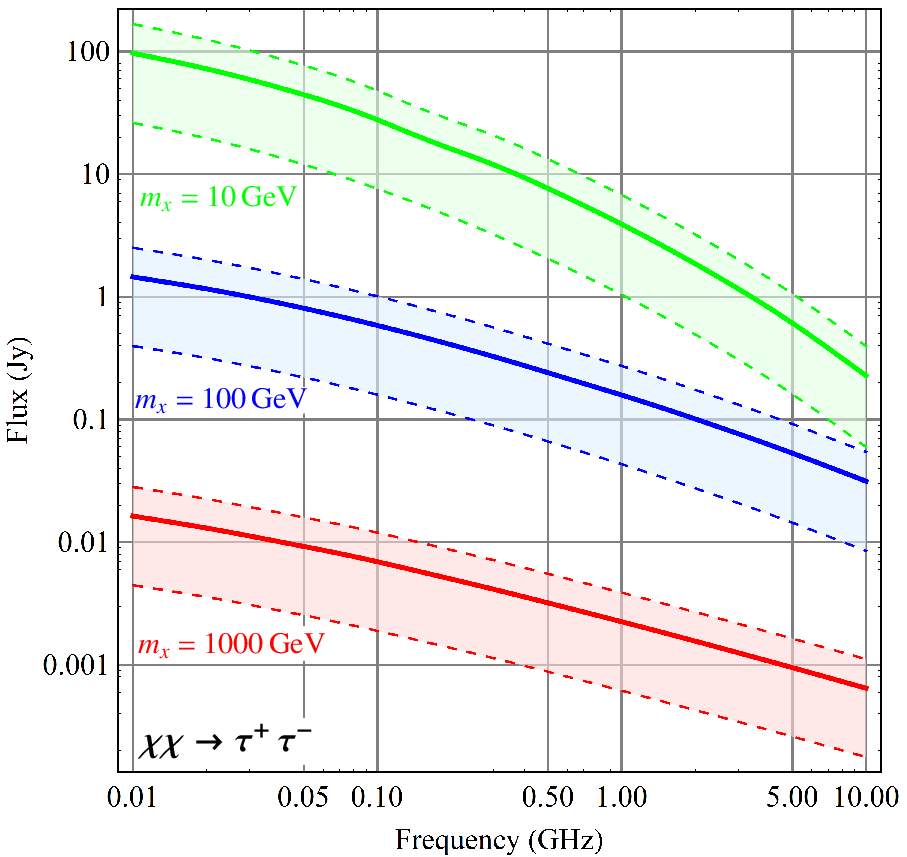} \\ }
\end{minipage}
\caption{The frequency dependence of the expected radio flux for three different WIMP masses and two annihilation channels. ROI radius $\alpha = 5'$, shaded regions reflect uncertainty in DM density distribution as on the previous plots, $B(0,0)=50~\mu$G, $\left\langle \sigma v\right\rangle = 3 \cdot 10^{-26}$ cm$^3$/s. More details are in subsection \ref{Estimated fluxes}.}
\label{S(nu)}
\end{figure*}

Another important result of our computations is the frequency dependence of radio flux from M31 halo. These dependencies are shown on fig. \ref{S(nu)} for three different WIMP masses, both annihilation channels, both limiting halo models and the chosen ROI with $\alpha = 5'$. As a general feature, we see, for all cases,  a decrease of the radio flux as the frequency increases. The steepness of this decrease is, however,  more pronounced for lower WIMP masses.
% and vice versa. 
This behavior suggests to primarily leverage on  low frequency observations for obtaining DM constraints. However, 
the sensitivity of radio surveys typically increases with frequency, partially compensating the effect of the lower signal.
% the flux decreases not very steeply, and the absolute sensitivity of radio surveys typically increases with frequency. Thus, these two effects would work against each other, and it is not obvious which frequencies are optimal for  inferring of the strongest constraints possible. 
There is, therefore, no obvious optimal frequency to be used and  that would lead to the strongest constraints possible.
 For this reason, we  opted  radio observations at  very different frequencies for actual constraints derivation. This procedure and its results will be described in the next section.

\section{Obtaining constraints on DM annihilation \label{Obtaining constraints on DM annihilation}}

\subsection{Comparison with radio surveys \label{Comparison with radio surveys}}
In order to make a next step on the way to final constraints, we needed to obtain the upper limits on the actually observed radio fluxes from our ROI. Comparison of these upper limits with the expected fluxes computed above immediately provides the upper limits on the annihilation cross section for different values of WIMP masses, since the expected fluxes is directly proportional to $\left\langle \sigma v\right\rangle$. For this purpose we studied many major radio surveys conducted in the past and chose four of them  (corresponding frequencies and detailed descriptions are provided in brackets): VLSS (74 MHz, \cite{VLSS}), WENSS (325 MHz, \cite{WENSS}), NVSS (1400 MHz, \cite{NVSS}) and GB6 (4850 MHz, \cite{GB6}). All other surveys were rejected due to either one of the following two reasons: too low resolution of a survey (i.e. the telescope beam size is larger than the ROI size) or not covering M31 area of the sky. 
%But the selected four surveys span very wide frequency range and provide significant constraints as we will see below. 
In  table \ref{table} we outlined the  main surveys' parameters, which were collected from \cite{VLSS}-\cite{GB6} and then used in calculations. The uncertainties of survey frequencies, caused by finite bandwidth, do not exceed the level of $\sim$ 4\%, and do not affect final results significantly.

\begin{table}
\caption{Main parameters of all radio surveys used. \label{table}}
 \centering
  \begin{tabular}{|m{0.22\linewidth}|m{0.22\linewidth}|m{0.22\linewidth}|m{0.22\linewidth}|}
\hline
\begin{center}
\textbf{Survey}
\end{center} & 
\begin{center}
\textbf{Frequency, MHz}
\end{center} & 
\begin{center}
\textbf{Beam diameter or FWHM $\mathbf{2 \beta_i}$, arcseconds}
\end{center} & 
\begin{center}
\textbf{RMS noise level inside the beam $\mathbf{\sigma_i^{(b)}}$, mJy}
\end{center} \\
\hline
\begin{center}
VLSS
\end{center} &
\begin{center}
74
\end{center} &
\begin{center}
80
\end{center} &
\begin{center}
100
\end{center}\\
\hline
\begin{center}
WENSS
\end{center} &
\begin{center}
325
\end{center}&
\begin{center}
82
\end{center}&
\begin{center}
4.0
\end{center}\\
\hline
\begin{center}
NVSS
\end{center} &
\begin{center}
1400
\end{center} &
\begin{center}
46
\end{center} &
\begin{center}
0.5
\end{center}\\
\hline
\begin{center}
GB6
\end{center} &
\begin{center}
4850
\end{center}&
\begin{center}
240
\end{center}&
\begin{center}
4.0
\end{center}\\
\hline
  \end{tabular}
 \end{table}

\begin{figure*}
\begin{minipage}[h]{0.49\linewidth}
\center{\includegraphics[width=1\linewidth]{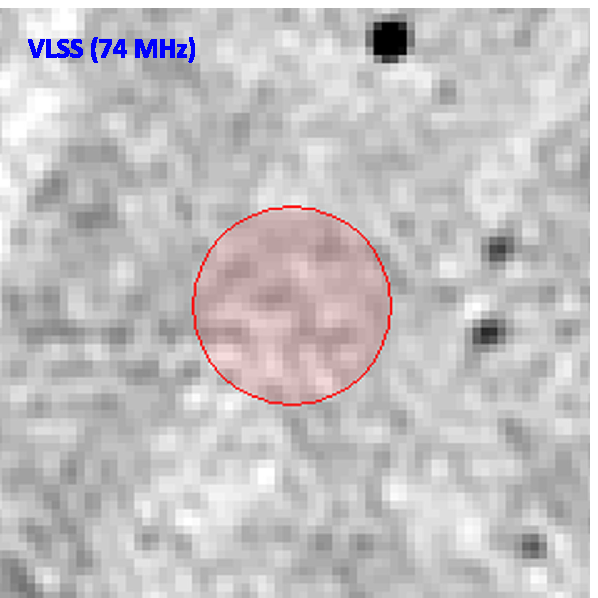}}  \\
\end{minipage}
\hfill
\begin{minipage}[h]{0.49\linewidth}
\center{\includegraphics[width=1\linewidth]{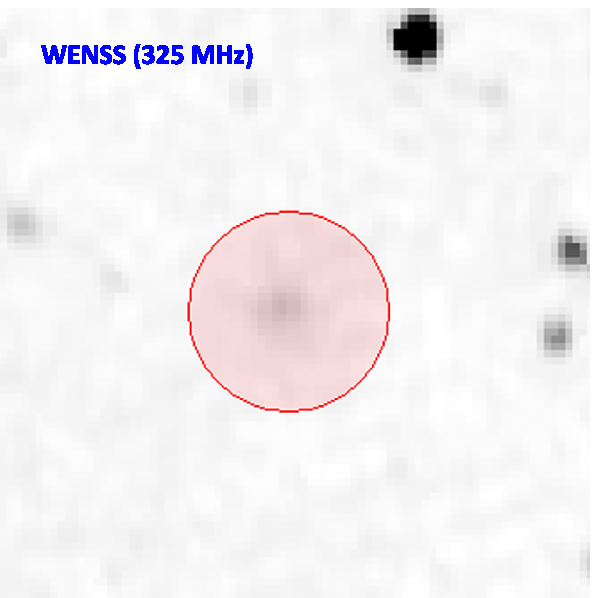}} \\
\end{minipage}  
\vfill   
\begin{minipage}[h]{0.49\linewidth}
\center{\includegraphics[width=1\linewidth]{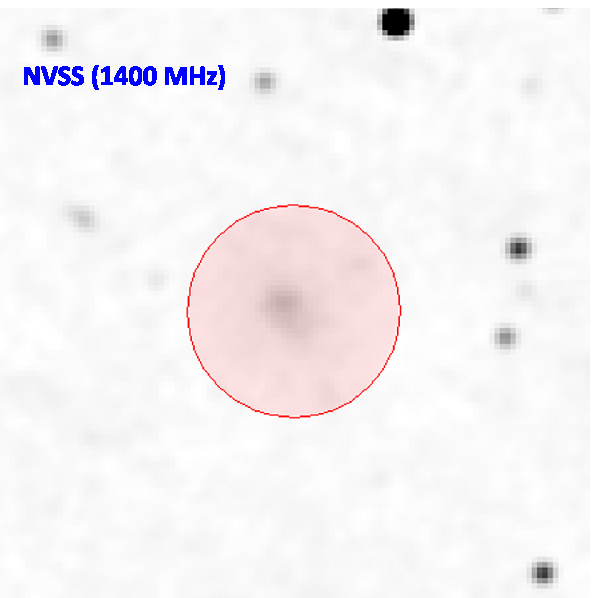}}  \\
\end{minipage}
\hfill
\begin{minipage}[h]{0.49\linewidth}
\center{\includegraphics[width=1\linewidth]{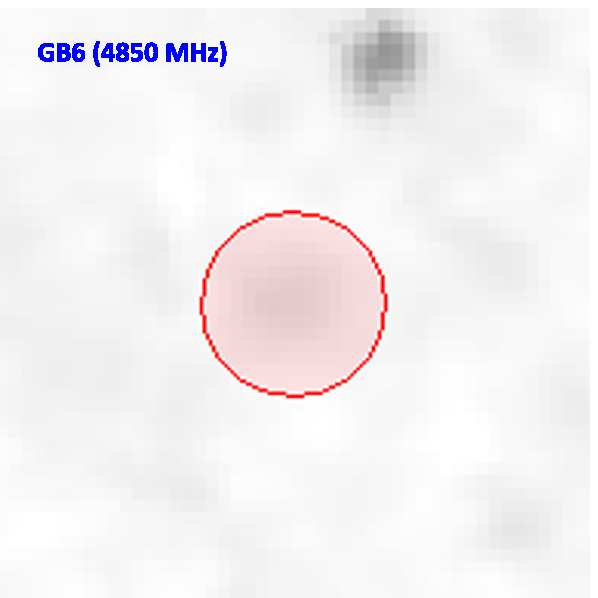}} \\
\end{minipage}
\caption{Radio images of the M31 central region. Our ROI of $5'$ radius is marked by the red circle. Corresponding surveys are commented on each image. We can find noise only on the 74 MHz and significant signals on the other frequencies.}
\label{Maps}
\end{figure*}

Real sky images were viewed and analysed by the Aladin software, which is described in \cite{Aladin}. The images of our ROI from all four radio surveys are shown on fig. \ref{Maps}. Firstly, we notice  that our ROI does not capture any contaminating point sources.
On the last three images (325, 1400 and 4850 MHz) some signal from M31 center is clearly visible. The 74 MHz image presents essentially noise. For the precise flux upper limit derivations we needed to obtain the measured signal values $c_i$ ($i=1-4$ denotes different observational frequencies) and the rms (root of mean square or root of dispersion) noise values $\sigma_{i}$ for our ROI for all images (we assume Gaussian distributions of noise levels inside a radio telescope beam and inside the ROI). The measured signal values were easily read by the Aladin. As for the rms noise values, we derived from basic principles that they can be estimated as $\sigma_{i} = \sigma_i^{(b)} \sqrt{N_i} = \sigma_i^{(b)} \frac{\alpha}{\beta_i}$, where $N_i$ are the numbers of beams contained in the ROI, $\beta_i$ are the angular radii of beams (or half of the FWHM for each survey) and $\sigma_i^{(b)}$ are the rms noises inside a beam for each survey (cited in table \ref{table}). Thus, we have obtained both necessary ingredients for a flux upper limit estimation from the ROI - the measured signal values and the rms noise values. 

Having all the necessary information, we obtained the limiting cross section values for cases of different parameter sets using the following formula for the probability density of the noise values inside our ROI:
\begin{equation}\label{p(n)}
	p_i(n_i = c_i-w_i(\left\langle \sigma v\right\rangle)) = \frac{1}{\sqrt{2 \pi} \sigma_i} \exp\left(-\frac{(c_i-w_i(\left\langle \sigma v\right\rangle))^2}{2 \sigma_i^2}\right),
\end{equation}
where $n_i$ - the noise values inside the ROI; $c_i$ - the measured signal values from the ROI and $w_i(\left\langle \sigma v\right\rangle)$ - the expected signals from WIMP annihilation, which is linearly proportional to an unknown annihilation cross section $\left\langle \sigma v\right\rangle$. Here we do not include any background radiation (this question will be further developed in the next section). For our final constraints onto $\left\langle \sigma v\right\rangle$ we decided to use 99.73\% confedence level, which corresponds to $3\sigma$ Gaussian confedence level.

\begin{figure*}
\begin{minipage}[h]{0.49\linewidth}
\center{\includegraphics[width=1\linewidth]{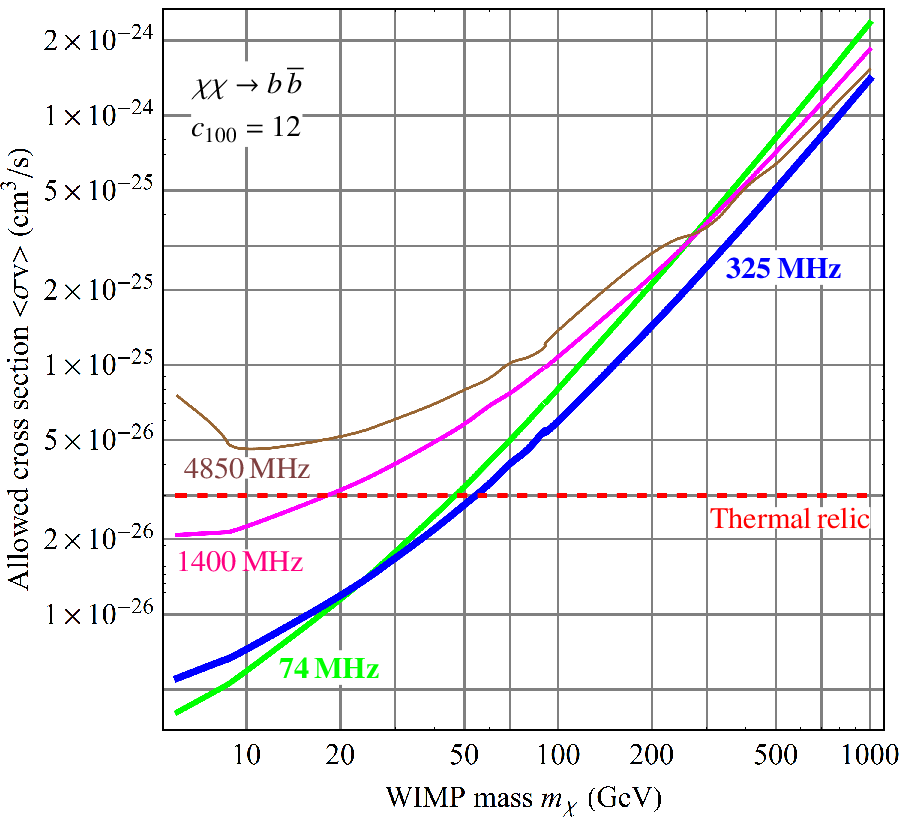}}  \\
\end{minipage}
\hfill
\begin{minipage}[h]{0.49\linewidth}
\center{\includegraphics[width=1\linewidth]{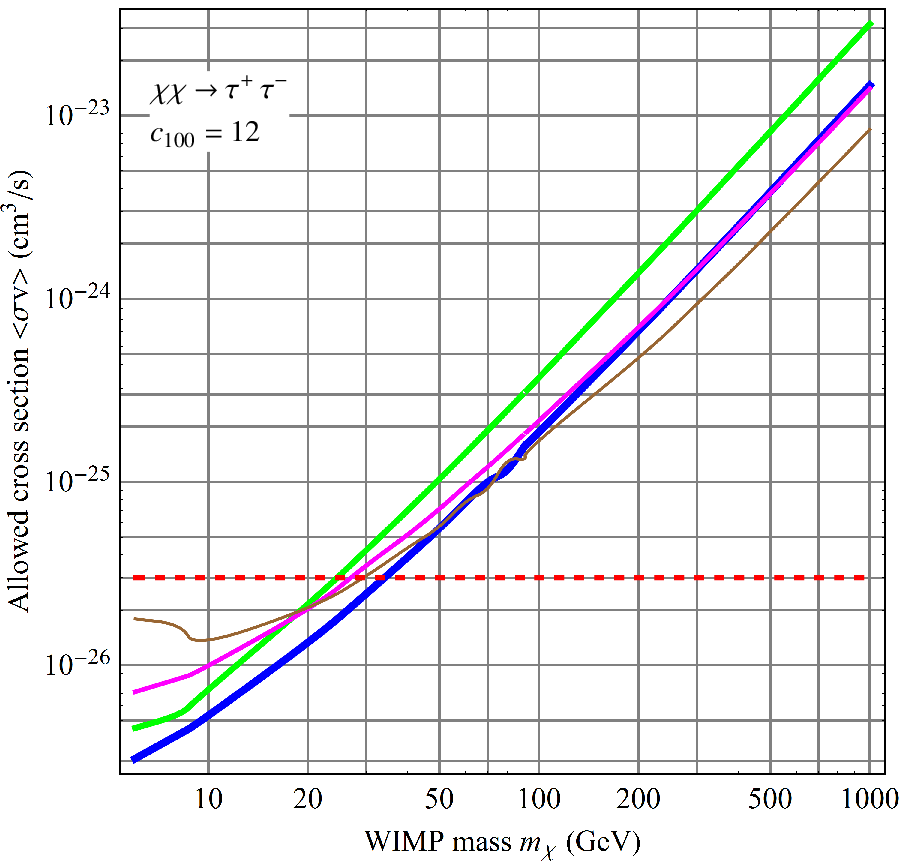}} \\
\end{minipage}
\vfill
\begin{minipage}[h]{0.49\linewidth}
\center{\includegraphics[width=1\linewidth]{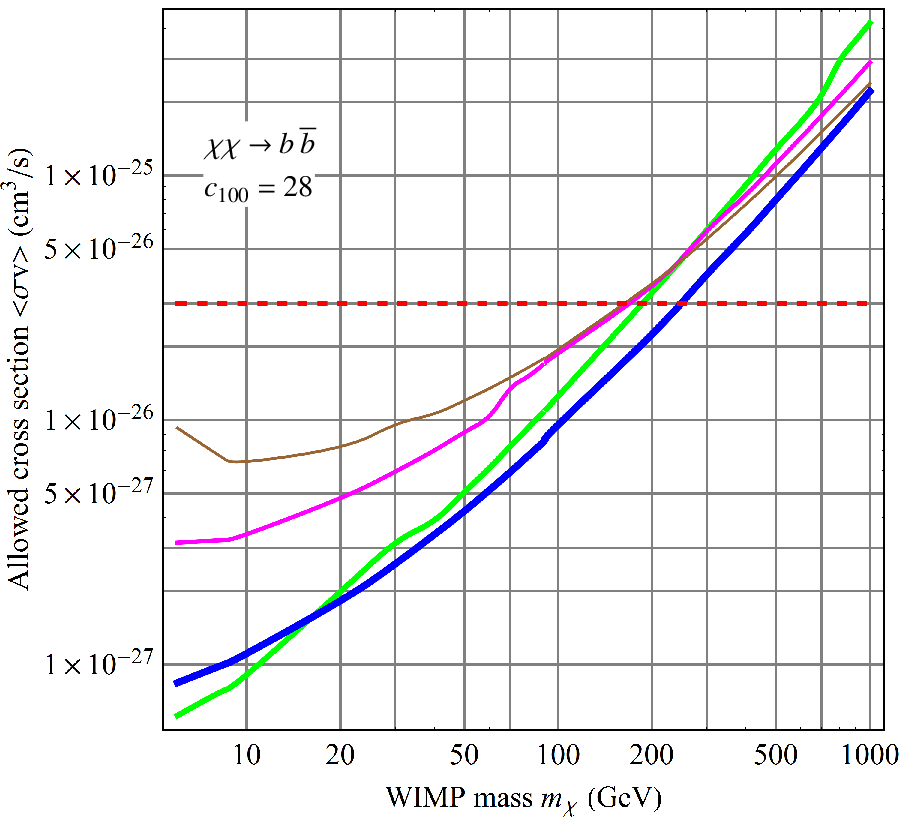}}  \\
\end{minipage}
\hfill
\begin{minipage}[h]{0.49\linewidth}
\center{\includegraphics[width=1\linewidth]{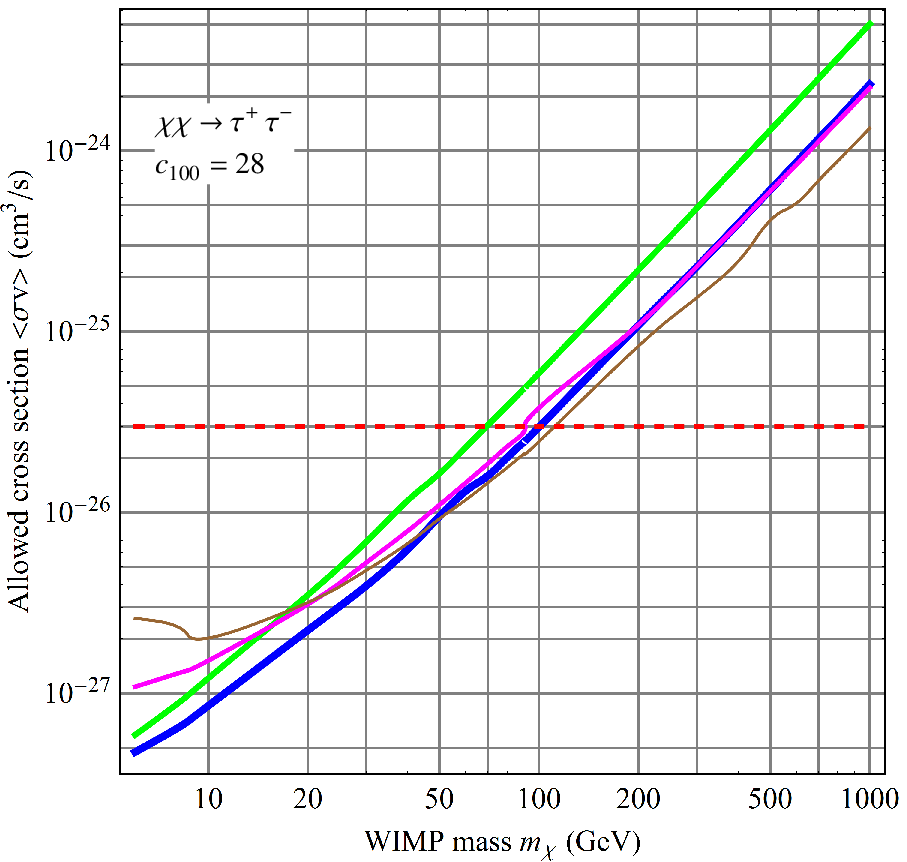}} \\
\end{minipage}
\caption{All the computed constraints: the WIMP annihilation cross section values above the contours are excluded at 99.7\% confedence level. Constraints from different surveys are marked by different colors and thicknesses, which are explained on one of the plots and same for all other plots. Standard thermal relic value $\left\langle \sigma v\right\rangle = 3 \cdot 10^{-26}$ cm$^3$/s is also shown by the red dashed line. Corresponding halo model and annihilation channel are commented on each plot. No emission sources besides DM are assumed. Magnetic field has the most probable distribution with $B(0,0)=50~\mu$G. For more details see subsection \ref{Comparison with radio surveys}.}
\label{FC}
\end{figure*}

In order to guide intuition on which experiment is the most constraining for a given particle mass, 
in fig. \ref{FC}  we showed our results on the cross section when each frequency is considered individually.
There we plotted the constraints obtained at 99.7\% confidence level by four different surveys used, for two halo models (described in subsection \ref{DM density distribution}), two annihilation channels selected $\chi\chi \rightarrow b\bar{b}, \tau^+\tau^-$ and the most probable magnetic field distribution with $B(0,0)=50~\mu$G. We spanned the range of WIMP masses between 6 GeV and 1000 GeV and computed the limiting $\left\langle \sigma v\right\rangle$ values  for a subset of WIMP masses inside this range, interpolating results for mass values  in between. %Then obtained sets of points were joined into smooth curves by second-order interpolation.
The general key  properties of our results can be summarized as follows:
 first of all, different halo models yield very different  limiting annihilation cross sections (by about one order of magnitude) for all WIMP masses. This reflects the  high sensitivity of our results to the halo model assumed. Another noticeable  fact is that $\tau^+\tau^-$ annihilation channel appears to be less constrained than $b\bar{b}$ channel,  in complete agreement with the mentioned expectation that harder secondary spectra channels produce generally lower final radiation flux.
 It is also interesting to notice that,
  %Also we see, that 
  for different WIMP masses and annihilation channels, the most constraining survey among all the ones considered varies. 
  %different surveys puts the strongest constraints  
  Therefore, we can not isolate  a specific key frequency/experiment for each plot,
   %which would be 
 bearing the role of   the most constraining for all WIMP masses - we need to use all experiments  together to get the best constraints. And final constraints from the joint analysis will be gathered in the next section together with analysis of the role of magnetic field uncertainties.

\subsection{Joint analysis and final constraints \label{Joint analysis}}
Constraints shown on fig. \ref{FC} present exclusion contours from four independent observations. As we can see, these contours lie rather close to each other on each diagram at least for some WIMP masses. Such situation suggests to apply a joint likelihood analysis in order to infer combined constraints from all four independent measurements, which we can expect to be better than the constraints taken  just considering the  exclusion curves in fig. \ref{FC}. 

\begin{figure*}
\begin{minipage}[h]{0.49\linewidth}
\center{\includegraphics[width=1\linewidth]{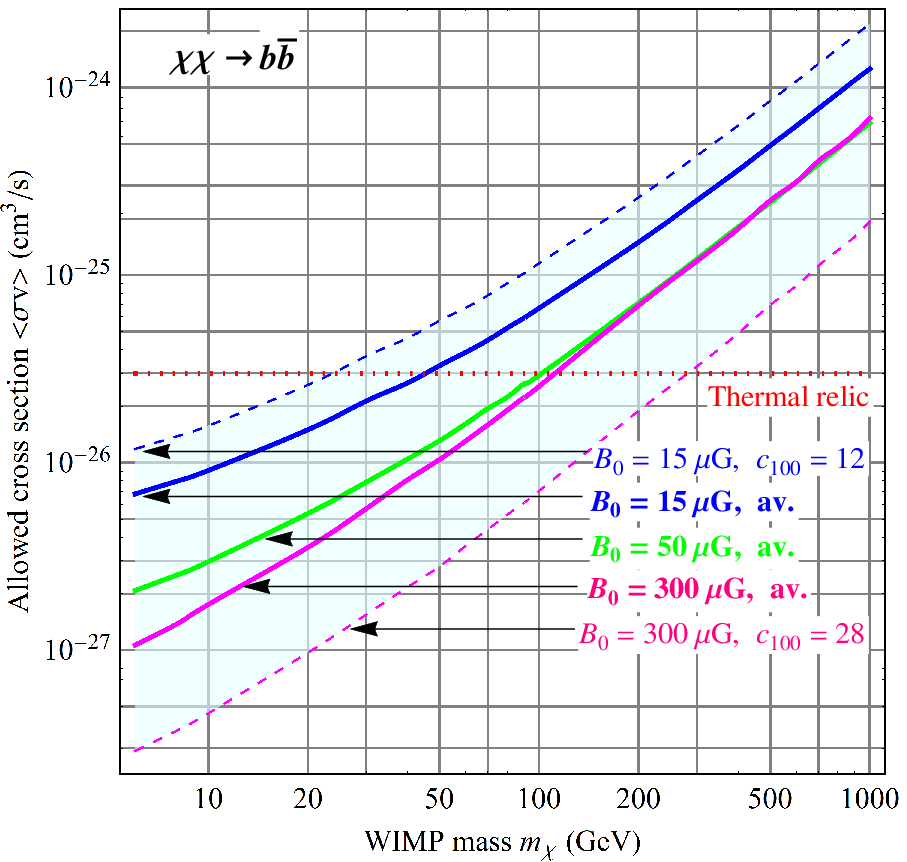} \\ }
\end{minipage}
\hfill
\begin{minipage}[h]{0.49\linewidth}
\center{\includegraphics[width=1\linewidth]{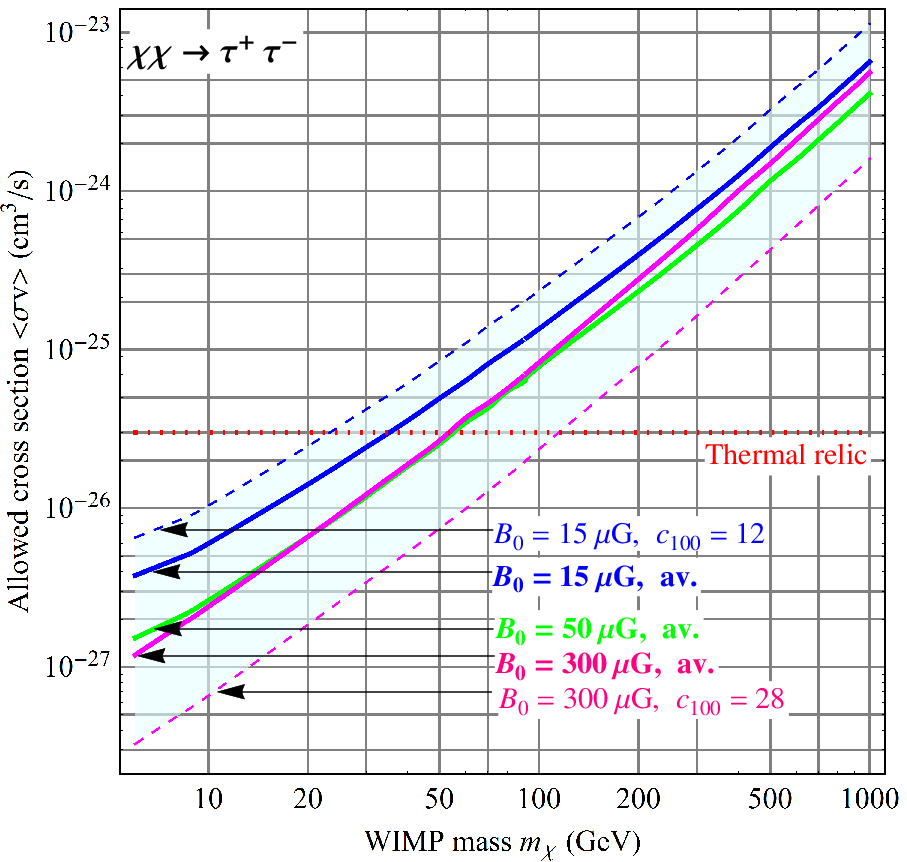} \\ }
\end{minipage}
\caption{Joint constraints including all radio surveys and backgrounds. These plots show all potential uncertainties of both distributions of DM density and magnetic field strength. The dashed contours reflect specific halo models discussed in the text (lines represent 99.7\% confedence level). The continuous lines present averages between the exclusions for the corresponding limiting halo models. Different continuous contours illustrate the variation of exclusions due to magnetic field uncertainties. The upper and lower exclusions enclose the shaded regions of all other possible exclusions from the most conservative to the most optimistic ones. More details are in subsection \ref{Joint analysis}.}
\label{All-1}
\end{figure*}

\begin{figure*}
\begin{minipage}[h]{0.49\linewidth}
\center{\includegraphics[width=1\linewidth]{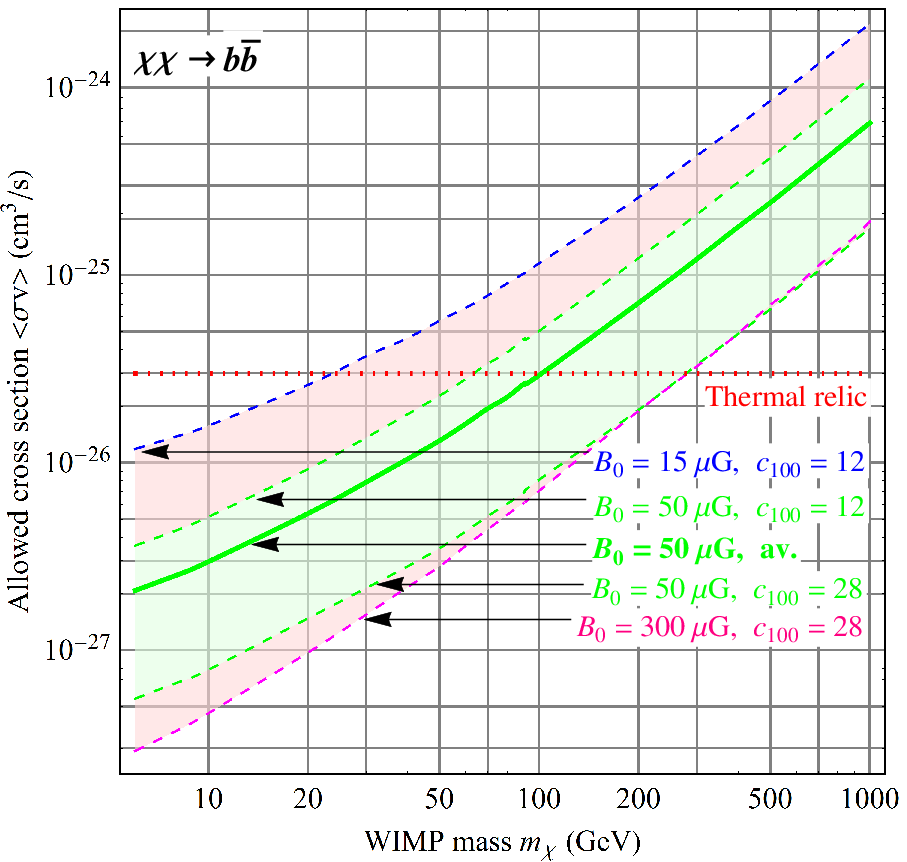} \\ }
\end{minipage}
\hfill
\begin{minipage}[h]{0.49\linewidth}
\center{\includegraphics[width=1\linewidth]{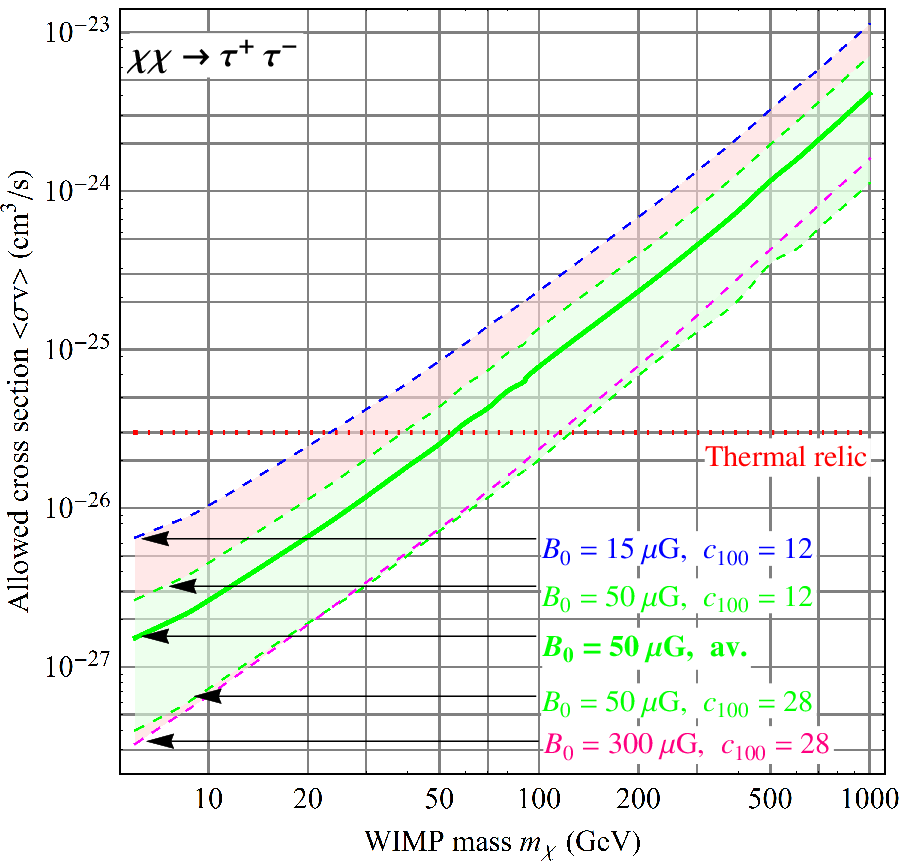} \\ }
\end{minipage}
\caption{Joint constraints obtained. These plots incorporate potential uncertainties of both distributions of DM density and magnetic field strength. The continuous line presents the most realistic exclusion contour, which corresponds to the result of our work on the fiducial case we assumed (average exclusion between those which correspond to the two limiting halo cases and the most realistic magnetic field distribution with $B(0,0)=50~\mu$G). The shaded area around the continuous contour, which is constrained by the inner dashed green lines, corresponds to dark halo model uncertainties only. Larger shaded region, which is enclosed into the outer dashed lines, incorporates both uncertainties of dark halo and magnetic field. For more details see subsection \ref{Joint analysis}.}
\label{All-2}
\end{figure*}

Signals other than DM are expected at each frequency. An appropriate modeling of all other astrophysical signals, with their frequency dependence, is necessary  in the joint analysis of various frequency maps.
We  decided to  make no  assumptions on  astrophysical processes other than DM annihilation, and simply modeled the total contribution of all other background signals as an a--priori unconstrained offset uncorrelated between frequency bands.
We then performed a standard Bayesian analysis of our data. Following the Bayes theorem  (for more details see \cite{Bayes}):
\begin{equation}\label{5D-density}
p(\vec{\Theta}|\vec{c}) = \frac{p(\vec{c}|\vec{\Theta})p(\vec{\Theta})}{p(\vec{c})},	
\end{equation}
where the l.h.s. is the desirable probability density distribution for unknown set of parameters $\vec{\Theta}$ in case of the observation of data $\vec{c}$, $p(\vec{c}|\vec{\Theta}) \equiv L(\vec{\Theta})$ is the probability density of our observed data $\vec{c}$ for fixed parameters $\vec{\Theta}$ (or likelihood function), $p(\vec{\Theta})$ is the prior probability density of $\vec{\Theta}$ and $p(\vec{c})$ is the marginal likelihood. In our case $\vec{\Theta} = (\left\langle \sigma v\right\rangle, s_i)$ is the set of unknown parameters, which we are aiming to infer - annihilation cross section and background radiation fluxes $s_i$  on all four frequencies involved. $\vec{c} = (c_i)$ is the data measured. Then the likelihood function $L$ will have the form:
\begin{equation}\label{L}
	p(\vec{c}|\vec{\Theta}) = L(\vec{\Theta}) \sim \prod_{i=1}^4 \exp\left(-\frac{(c_i-s_i-w_i(\left\langle \sigma v\right\rangle))^2}{2\sigma_i^2}\right).
\end{equation}
This likelihood function presents essentially the product of the independent noise level Gaussian distributions on all frequencies written in terms of the measured and expected fluxes (taking into account that $c_i = s_i+w_i+n_i$). As for $p(\vec{\Theta})$, we made no specific assumptions and chose a constant (flat) prior.
The function $p(\vec{c})$ in eq. (\ref{5D-density}) does not require specification, since it does not depend on $\vec{\Theta}$ and, hence, plays a role of a normalization constant. After we constructed the five-dimensional density distribution (\ref{5D-density}) for all five unknown quantities, we needed to integrate eq. (\ref{5D-density}) over all possible ranges of non-interesting parameters (marginalization). In our case we were not interested in background values, so we obtained the distribution for the desired quantity $\left\langle \sigma v\right\rangle$ by marginalizing over the background values:  
\begin{multline}\label{p(w)}
	p(\left\langle \sigma v\right\rangle|\vec{c}) \sim \int L(\vec{\Theta})p(\vec{\Theta})d\vec{s} \\ = \iiiint\limits_0^{\infty} \prod_{i=1}^4 \exp\left(-\frac{(c_i-s_i-w_i(\left\langle \sigma v\right\rangle))^2}{2\sigma_i^2}\right)ds_1..ds_4, 
\end{multline}
where we took into account the flat prior. Thus formula (\ref{p(w)}) gives us the final probability density distribution for annihilation cross section combined from all four observations with possible unknown backgrounds included. From this distribution we easily constructed the final exclusion contours. We presented them for the case of 99.73\% confidence level on fig. \ref{All-1}-\ref{All-2}, where the allowed area is below the exclusion contours for the corresponding models.

We included in our final results possible uncertainties in the magnetic field distribution discussed in subsection \ref{Magnetic field}. 
As we outlined there, this distribution is uncertain in two aspects - its vertical scale height $z_0'$ and the central field strength $B(0,0)$. As for the $z_0'$, our trial runs showed that $z_0'$ variation over all possible values is able to change the final exclusion values of $\left\langle \sigma v\right\rangle$ by no more than $\left\langle \sigma v\right\rangle^{+5\%}_{-20\%}$ for all possible sets of model parameters. 
%Considering an overall smallness of such variation (in comparison with that one associated with halo model uncertainties) and its tendency towards stronger constraints we did not include $z_0'$ variation into the final results. 
This variation is considerably smaller than the one induced by, e.g. halo model uncertainties; and would  suggest less conservative limits. For these reasons, we do not consider it in our final results. 
However, we can not treat in this way the second important parameter - $B(0,0)$, because practice showed significant variation of the final exclusions due to uncertainties in $B(0,0)$ discussed in subsection \ref{Magnetic field}. According to this discussion, we expect $B(0,0) = 50~\mu$G to be the most probable, but we also consider here   $B(0,0)$   values of $\sim 15$ and $\sim 300~\mu$G as limiting cases. We obtained all our final exclusions for these three different magnetic field distributions.
 
Our results are presented in  fig. \ref{All-1}-\ref{All-2}. 
In these figures, the dashed lines represent the specific halo models (either $c_{100}=12$ or $c_{100}=28$), while the solid lines correspond to the algebraic averages between  the two  limiting halo models, for fixed magnetic field values.
The two figures show limiting curves for different combinations of the  (three) possible magnetic field models and (two) possible halo models.
Fig. \ref{All-1} shows more explicitly the dependence on the field strength, while fig. \ref{All-2} shows the dependence on the halo model for the reference field strength $B(0,0) = 50~\mu$G, and allows to compare these uncertainties with the ones caused by variation in the magnetic field.

As we can see, annihilation with the standard thermal relic cross section $\left\langle \sigma v\right\rangle = 3 \cdot 10^{-26}$ cm$^3$/s is excluded for a significant part of WIMP mass range. We listed in table \ref{table 2} the precise WIMP mass values for all models considered, below which $\left\langle \sigma v\right\rangle$ is less than $3 \cdot 10^{-26}$ cm$^3$/s. These masses were obtained as the contact points between the corresponding exclusions and the thermal relic threshold on fig. \ref{All-1}-\ref{All-2}. We can see a significant spread in limiting masses around those which correspond to our preferred model with $B(0,0) = 50~\mu$G and the averaged flux between two limiting halo cases. For this most realistic model the WIMP masses  which do not allow the standard relic cross section  are the ones smaller than \textbf{100 GeV} and \textbf{55 GeV} for $b\bar{b}$ and $\tau^+\tau^-$ channels respectively.
% We emphasize these limiting masses as the main final result of our study. As for other possible annihilation channels, according to our previous notices we can expect for their exclusion contours to be lying approximately somewhere between exclusions for $b\bar{b}$ and $\tau^+\tau^-$. 
%In general, we can say that the final exclusion limits obtained suggest significant and competitive constraints, shrinking further allowed WIMP parameter space. 
In general, these limits were derived with rather conservative assumptions. %are consistent and not too distant from others obtained with other techniques and observations.
Discussion of our results and comparison with other studies will be done in the next section.

\begin{table*}
\caption{WIMP masses in GeV, below which annihilation with the canonical cross section is excluded, for the different halo and magnetic field models. The most realistic cut off values are emphasized. \label{table 2}}
 \centering
  \begin{tabular}{|m{0.13\linewidth}|m{0.13\linewidth}|m{0.13\linewidth}|m{0.13\linewidth}|m{0.13\linewidth}|m{0.13\linewidth}|m{0.13\linewidth}|}
\hline
\begin{center}
\textbf{Central field $\mathbf{B(0,0)}$}
\end{center} & 
\begin{center}
$\boldsymbol{c_{100}=12,}$ $\boldsymbol{\chi\chi \rightarrow b\bar{b}}$
\end{center} & 
\begin{center}
\textbf{Averaged flux, $\boldsymbol{\chi\chi \rightarrow b\bar{b}}$}
\end{center} & 
\begin{center}
$\boldsymbol{c_{100}=28,}$ $\boldsymbol{\chi\chi \rightarrow b\bar{b}}$
\end{center} &
\begin{center}
$\boldsymbol{c_{100}=12,}$ $\boldsymbol{\chi\chi \rightarrow \tau^+\tau^-}$
\end{center} & 
\begin{center}
\textbf{Averaged flux, $\boldsymbol{\chi\chi \rightarrow \tau^+\tau^-}$}
\end{center} & 
\begin{center}
$\boldsymbol{c_{100}=28,}$ $\boldsymbol{\chi\chi \rightarrow \tau^+\tau^-}$
\end{center} \\
\hline
\begin{center}
$\boldsymbol{15~\mu\mbox{\textbf{G}}}$
\end{center} &
\begin{center}
23
\end{center} &
\begin{center}
45
\end{center} &
\begin{center}
160
\end{center} &
\begin{center}
23
\end{center} &
\begin{center}
35
\end{center} &
\begin{center}
88
\end{center} \\
\hline
\begin{center}
$\boldsymbol{50~\mu\mbox{\textbf{G}}}$
\end{center} &
\begin{center}
63
\end{center}&
\begin{center}
\textbf{\textit{100}}
\end{center}&
\begin{center}
280
\end{center}&
\begin{center}
39
\end{center}&
\begin{center}
\textbf{\textit{55}}
\end{center}&
\begin{center}
130
\end{center}\\
\hline
\begin{center}
$\boldsymbol{300~\mu\mbox{\textbf{G}}}$
\end{center} &
\begin{center}
75
\end{center} &
\begin{center}
110
\end{center} &
\begin{center}
280
\end{center} &
\begin{center}
38
\end{center} &
\begin{center}
53
\end{center} &
\begin{center}
110
\end{center} \\
\hline
  \end{tabular}
 \end{table*}

\section{Summary and discussion \label{Summary and discussion}}

In our work we aimed to obtain robust constraints on WIMP--mass/annihilation--cross--section plane by studying radio observations of M31. Annihilating DM in M31 halo produces highly energetic secondary leptons, which, in turn, emit synchrotron radiation due to Andromeda magnetic field.
% on radio frequencies. 
We computed the expected characteristics of this radiation. We first modeled the DM density distribution in M31, using the standard NFW density profile. Parameters of this distribution were found in the previous studies of the M31 halo. These studies appeared to have good agreement between each other in halo mass determination ($M_{100} \approx 10^{12}M_{\odot}$), but they differ significantly in the obtained halo concentration parameter - $c_{100} \approx 12$ vs. $c_{100} \approx 28$. That's why we treated these two cases separately and considered them as limiting cases of possible halo models for M31. Another important step of the calculations was magnetic field distribution modeling. For this purpose we used available magnetic field measurements in M31. Information, which is not available for the M31, was gathered by extrapolating MW magnetic field distribution properties and general properties of galactic magnetic fields as well. The idea to use MW magnetic field measurements was motivated by high similarity between the M31 and the MW, and, also, by the  much more detailed knowledge we have about the MW. We introduced a realistic field distribution, which is characterized by the 50 $\mu$G value of the central field strength and exponential decline both in vertical and radial-in-plane directions. Then we computed the secondary particle yields from WIMP annihilation using the results of DarkSUSY package. We showed that
neglecting the spatial diffusion of the injected leptons is a reasonable approximation.
%As for propagation of injected leptons, we neglected  their spatial diffusion, which was shown to be a reasonable approximation. 
We calculated final fluxes in optically transparent media approximation. As for the sky area chosen for putting the constraints, 
%it was demonstrated that 
we chose the circle with the angular radius $5'$ around the M31 center, and motivated our choice in subsection \ref{Estimated fluxes}.
 %is an optimal choice (see subsection \ref{Estimated fluxes}). 
% And this choice was used for all subsequent work.
 After obtaining the radio fluxes expected, we compared them with available radio observations of M31. For this purpose we used four radio surveys spanning wide range of frequencies: VLSS (74 MHz), WENSS (325 MHz), NVSS (1400 MHz) and GB6 (4850 MHz). We analyzed the data from these surveys and compared them with the theoretical expectations for radio fluxes. We then derived limits on the cross section $\left\langle \sigma v\right\rangle$ as a function of particle mass (ranging from 6 GeV to 1000 GeV). Our analysis is conservative in sense of the absence of any specific assumptions about a possible radiation other than from DM. The final results are presented in the section \ref{Obtaining constraints on DM annihilation} on fig. \ref{All-1}-\ref{All-2}. We also computed our results for the alternative less probable magnetic field models with the central field values of 15 and 300 $\mu$G in order to study the dependence of our final constraints on this model parameter. 

Our main result is the exclusion of the
% opportunity of
possibility of  WIMP annihilation with the thermal cross section $\left\langle \sigma v\right\rangle = 3 \cdot 10^{-26}$ cm$^3$/s or higher for WIMP masses below $\approx 100$ GeV ($\approx 55$ GeV )  for the  $b\bar{b}$ ( $\tau^+\tau^-$) annihilation channel. However, these estimations are affected by significant uncertainties. Taking into account potential uncertainties in DM density and magnetic field distributions, these limits could be as low as 23 GeV and as high as 280 GeV for the $b\bar{b}$ channel. For annihilation into $\tau^+\tau^-$ these limits respectively are 23 GeV and 130 GeV (see table \ref{table 2}).   

%In a frame of other similar studies, our results look rather competitive. 
Our results are comparable  with other findings on WIMP masses derived through other observations.
The best up-to-day constraints in indirect DM searches were obtained by joint analysis of MW satellites by Fermi gamma telescope. This work is presented e.g. in \cite{Fermi-dwarfs}, where the authors obtained the constraints by comparison of Fermi data with the expected gamma emission produced directly in WIMP annihilation into the same channels. They reported the lowest allowed WIMP masses 40 GeV and 19 GeV for $b\bar{b}$ and $\tau^+\tau^-$ channels respectively. 
When considering 95\% systematic errors on the DM distribution within the dwarves, their derived WIMP mass lower limit varies between 
19--240 GeV ($b\bar{b}$) and 13--80 GeV ($\tau^+\tau^-$).
%Their uncertainty ranges are 19--240 GeV and 13--80 GeV respectively. 
%We can see that 
%Our limits  of 100 GeV and 55 GeV overpass Fermi limits.
The constraints derived in this work, suggesting masses above 100 GeV and 55 GeV  in order to obtain a relic cross section, are within the range of uncertainties of current Fermi results. This suggests the relevance  of multi-frequency approach in indirect detection: all wavelength ranges may make valuable contributions into an overall progress. As for other studies dedicated to indirect searches on radio frequencies, we can compare our results with the similar studies for MW and M33 - \cite{Borriello-MW} and \cite{M33} respectively. Our constraints are significantly stronger than the last cited, since the exclusion contours obtained by the authors of \cite{Borriello-MW},\cite{M33} lie above the thermal relic value $\left\langle \sigma v\right\rangle = 3 \cdot 10^{-26}$ cm$^3$/s on the WIMP--mass/annihilation--cross--section plane for any possible WIMP masses.
 
We show here, for the first time, that it is possible to constrain the WIMP parameter space by radio observations of  M31. Specifically, while making conservative assumptions, we were able to exclude small WIMP masses for conventional, thermally produced DM. However, the range of higher masses of hundreds GeV is still absolutely allowed and unexplored. This paper presents only the first relatively simple step in a comprehensive analysis of M31, and  does not yet include all potential uncertainties and relevant effects. 
%However, these first results showed high demand of M31 for indirect detection, which is motivating us to proceed more extended analysis of this galaxy. 
This work may be expanded 
%In our future work we are planning to expand our analysis further 
in two main directions. Firstly, astrophysical uncertainties can  be treated with a more general approach:  including, for example, spatial diffusion of annihilation products and possible variations in magnetic field modeling.  
In addition, while encouraged by current finding,  the potential to study this object with other current and upcoming radio telescope may be explored.
\\

\begin{acknowledgments}
We would like to thank Wendy Lane - the Principal Investigator of the VLSS radio survey - for valuable discussions regarding observational aspects. Also we are grateful to Jennifer Siegal-Gaskins for generally useful feedback. EP acknowledges support from JPL-Planck subcontract 1290790. EP and AE were partially supported by NASA grant NNX07AH59G.
\end{acknowledgments}

%\begin{acknowledgments}
%\end{acknowledgments}

\bibliography{Biblio}


\appendix
\section{Absorption analysis \label{Absorption analysis}}

Here we discuss all potentially relevant mechanisms of absorption of radio emission during its propagation from the M31 center to the observer. We focused only on a potential absorption inside M31 and did not consider possible absorption in MW, because the Andromeda galaxy lies far enough from the Galactic plane, and we do not expect any significant absorption in this direction on the sky. In general, inside the Andromeda there are two main agents of interstellar medium relevant for our purposes: a dust and an ionized gas. Among them a dust can be excluded as an absorber rather easily: typically a dust grain size does not exceed $~10^{-4}$ cm (see e.g. \cite{dust}). And the wavelengths of radio emission involved are $~10-1000$ cm, which is many orders of magnitude larger than grain sizes. In such case, when a wavelength is much greater than an obstacle size, it is well known that radiation does not interact with obstacles. Thus, we can conclude that a dust does not affect radiation in the considered frequency range.

A situation with the absorption by interstellar plasma is more tricky. Here we can distinguish several possible ways of absorption. First of all, let's check the Langmuir frequency of plasma involved. As well known, a radiation can not propagate through plasma, if its frequency is less than the Langmuir frequency of propagation medium. The Langmuir frequency is defined as
\begin{equation}\label{pl. freq.}
  \nu_{pl} = \frac{\omega_{pl}}{2\pi} = \sqrt{\frac{e^2 n}{\pi m_e}},
\end{equation}
where $n$ is the concentration of plasma electrons, which needs to be estimated. For this purpose we used the results of \cite{MW-gas}, where the gas density distribution was obtained for the central region of MW. Taking into account that MW and M31 are very similar galaxies, we can extrapolate results for MW on M31 staying at necessary level of accuracy. From \cite{MW-gas} we can easily see, that the plasma concentration in a galactic center can not exceed $n \lesssim 10$ cm$^{-3}$ in the worst case, which yields $\nu_{pl} \lesssim 30$ kHz. Thus, the Langmuir frequency is much smaller than the observational frequencies of MHz-GHz, which means that the interstellar plasma is transparent with respect to this absorption mechanism.

Another potentially relevant mechanism is a synchrotron self-absorption. Synchrotron emission, generated by relativistic leptons from DM annihilation, can be absorbed by neighbour emitting leptons. In order to estimate the level of this absorption quantitatively we computed the corresponding optical depth along our line of sight, which goes through the M31 center:
\begin{equation}\label{tau ss}
  \tau_{ss} = \int\limits_{los} \alpha_{ss} dl,
\end{equation}
where $\alpha_{ss}$ is the synchrotron self-absorption coefficient. We used the derived expression for $\alpha_{ss}$ from \cite{Rybicki} (formula (6.50) there):
\begin{multline}\label{alpha ss}
  \alpha_{ss} = -\frac{c^2}{8\pi\nu^2}\int\limits_{m_e c^2}^{m_{\chi} c^2}dE P_e(E,\nu,\vec{r}) \\ \cdot E^2 \frac{\partial}{\partial E} \left(E^{-2} \cdot 2\frac{dn_e}{dE}(E,\vec{r})\right),
\end{multline}
 where $P_e(E,\nu,\vec{r})$ is the synchrotron power of one lepton defined by eq. (\ref{P})-(\ref{P content}), $\frac{dn_e}{dE}(E,\vec{r})$ is the stationary energy spectrum of emitting leptons of one kind derived by eq. (\ref{dn/dE}). Then we substituted all relevant values of the parameters involved. Particularly, for WIMP mass and frequency of observations we used as an example $m_{\chi} = 100$ GeV and $\nu = 74$ MHz. Such combination of all parameters yields $\tau_{ss} \sim 10^{-3}$, which means that we can neglect by synchrotron self-absorption completely and reliably. And this conclusion is valid for all other frequencies used in our work, because $\alpha_{ss}$ decreases with frequency, and 74 MHz is the lowest frequency of observations. In general, this conclusion agrees with the results of \cite{MW-ss}, where the authors solved similar DM annihilation problem with application to MW. And they also showed an irrelevance of synchrotron self-absorption on all frequencies.

And the last potential absorption mechanism, which has to be checked, is a free-free absorption by plasma. Here we also estimated the corresponding optical depth:
 \begin{equation}\label{tau ss}
  \tau_{ff} = \int\limits_{los} \alpha_{ff} dl,
\end{equation}
where the absorption coefficient $\alpha_{ff}$ was also taken from \cite{Rybicki} (formula 5.18b there):
\begin{equation}\label{alpha ff}
 \alpha_{ff} = 3.7 \cdot 10^8 T^{-1/2} Z^2 n^2 \nu^{-3} \left(1-\exp\left(-\frac{h\nu}{kT}\right)\right)g_{ff},
\end{equation}
where $n$ and $T$ are plasma concentration and temperature respectively. As for the spatial distribution $n(\vec{r})$, it does not appear to be obtained in the literature for M31. That's why we decided to use such distribution for MW again as an approximation. Taking all relevant information from \cite{MW-gas} and assuming $g_{ff} \approx 1$ and $Z = 1$ (hydrogen plasma) we estimated $\tau_{ff} \lesssim 0.01$ in the worst case scenario. Thus, we can see that the free-free absorption is not relevant for our work as well in the first approximation.

Summarizing, in this section we have conducted the detailed analysis of all potentially relevant absorption mechanisms of radio emission generated by DM annihilation. None of these mechanisms achieves significant level. This conclusion is in agreement with similar studies \cite{MW-ss}, \cite{Borriello-MW} for the MW. Thus we can ignore any absorption in our analysis without significant loss of accuracy.

\section{Spatial diffusion of annihilation products \label{Spatial diffusion of annihilation products}}

\begin{figure}[b]
  \centering
  \includegraphics{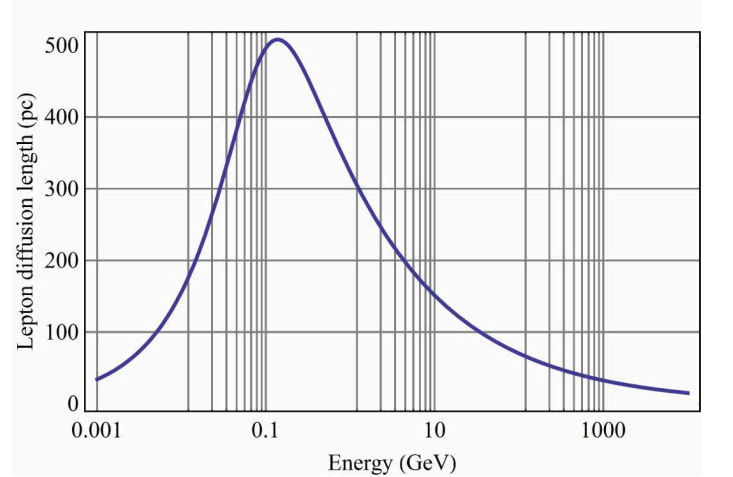}\\
  \caption{The dependence of the lepton diffusion length $l_D$ on the energy $E$. The magnetic field used is $B = 50~\mu$G, concentration is $n = 0.1$ cm$^{-3}$.}\label{plot diffusion}
\end{figure}

Here we study the role of the spatial diffusion of annihilation products in our problem. In order to understand the importance of the spatial diffusion for final results, we should compare the characteristic distance, which annihilation products travel while they are emitting relevant radiation, with the characteristic size of the emitting region. According to e.g. \cite{Borriello-MW}, the diffusion path traveled by leptons can be calculated as $l_D = \sqrt{D(E) \tau_{loss}(E)}$, where $D(E)$ is the diffusion coefficient for leptons participating in eq. (\ref{diffusion}); $\tau_{loss}(E)$ is the cooling time of leptons, during which they are emitting the expected radiation and losing their energy until escaping the relevant energy domain. We took the diffusion coefficient $D(E)$ from \cite{Borriello-MW}, where it is provided for the MW, and expected to work roughly for the Andromeda galaxy as well:
\begin{equation}\label{D}
  D(E) = D_0\left(\frac{E}{E_0}\right)^{\delta},
\end{equation}
with $D_0 = 10^{28}$ cm$^2$/s, $E_0 = 3$ GeV and a Kolmogorov spectrum $\delta = 1/3$. The cooling time for leptons can be estimated as
\begin{equation}\label{tau loss}
  \tau_{loss}(E) \sim \frac{E}{\dot{E}} = \frac{E}{b(E,\vec{r})},
\end{equation}
where $b(E,\vec{r}) = b_{ICS}+b_{sync}+b_{brem}+b_{Col}$ (see eq. (\ref{b_ICS})-(\ref{b_Col})). After all necessary substitutions we obtained $l_D$ dependence on the lepton energy $E$, which is shown on fig. \ref{plot diffusion}. For the magnetic field $B$ and the concentration $n$ we used the expected values for the M31 center $\sim 50 ~\mu$G and $\sim 0.1$ cm$^{-3}$ respectively. These parameter values were justified in the section \ref{Computing the radio flux}. As we can see on fig. \ref{plot diffusion}, the lepton diffusion path over relevant range of energies does not exceed $\sim$ 500 pc. Our emitting region, which we capture by our ROI with angular radius $\alpha \approx 5'$, would have the form of the cylinder with the radius $\rho_{max} \approx \alpha d \approx 1100$ pc. Thus, the smallest size of the emitting region is about 2 times larger than the diffusion length of leptons in the M31 center. It means, in turn, that the leptons do not have enough time to migrate significantly and escape from the emitting volume, before they cool down and discontinue radiation. Also taking into account the fact that the majority of total radiation flux due to DM annihilation is formed in the very central region, we can conclude that inclusion of the spatial diffusion in our calculations should not affect our results significantly. And neglecting by the diffusion is an acceptable approximation in our computation procedure. Such conclusion is in general agreement with the work \cite{Borriello-MW}, where the similar procedure of constraints derivation was conducted for the MW. However, since the spatial scale of the emitting volume and the lepton diffusion path do not differ drastically, we allow an opportunity to include the diffusion in our calculations in a future work in order to improve accuracy of our results.

\end{document}